\documentclass[a4paper,11pt]{article}

\usepackage{a4wide}
\usepackage[english]{babel}

\usepackage{amsfonts}
\usepackage{amsmath}
\usepackage{graphicx}
\usepackage{caption}
\usepackage[colorlinks,bookmarks=true]{hyperref}
\usepackage{amsthm}
\usepackage{enumitem}

\newtheorem{lemma}{Lemma}[section]
\newtheorem{definition}{Definition}[section]
\newtheorem{theorem}{Theorem}[section]
\newtheorem{prop}{Proposition}[section]
\newtheorem{remark}{Remark}[section]

\newtheorem{assumption}{Assumption}[section]
\newtheorem{rem}{Remark}[section]

\newtheorem{corollary}{Corollary}[section]

\newcommand{\E}{\mathbb{E}}

\title{Perfect hedging in rough Heston models}

\author{Omar El Euch\\ \'Ecole Polytechnique \\ omar.el-euch@polytechnique.edu\\$~~$\\
Mathieu Rosenbaum\\ \'Ecole Polytechnique\\
  mathieu.rosenbaum@polytechnique.edu}

\begin{document}

\maketitle

\begin{abstract}
\noindent Rough volatility models are known to reproduce the behavior of historical volatility data while at the same time fitting the volatility surface remarkably well, with very few parameters. However, managing the risks of derivatives under rough volatility can be intricate since the dynamics involve fractional Brownian motion. We show in this paper that surprisingly enough, explicit hedging strategies can be obtained in the case of rough Heston models. The replicating portfolios contain the underlying asset and the forward variance curve, and lead to perfect hedging (at least theoretically). From a probabilistic point of view, our study enables us to disentangle the infinite-dimensional Markovian structure associated to rough volatility models.  
\end{abstract}


\section{Introduction}\label{intro}

It has been recently shown in \cite{gatheral2014volatility} that rough fractional processes enable us to reproduce very accurately the behavior of historical volatility time-series. More precisely, the dynamic of their logarithm is quite similar to that of a fractional Brownian motion with Hurst parameter of order $0.1$. Recall that a fractional Brownian motion $W^H$ with Hurst parameter $H\in(0,1)$ can be built from a classical two-sided Brownian motion $W$ through the Mandelbrot-van Ness representation:
\begin{equation*}
W_t^H = \frac{1}{\Gamma(H+1/2)} \int_{-\infty}^0 \big((t-s)^{H-\frac{1}{2}} - (-s)^{H-\frac{1}{2}}\big)dW_s + \frac{1}{\Gamma(H+1/2)} \int_0^t (t-s)^{H-\frac{1}{2}} dW_s.
\end{equation*}
The fractional Brownian motion has H\"older regularity $H-\varepsilon$ for any $\varepsilon > 0$. Hence fractional volatility models with small Hurst parameter are referred to as rough volatility models.\\

\noindent Beyond historical data modeling, rough volatility models provide excellent fits and dynamics for the whole volatility surface, in particular for the at-the-money skew, with very few scalar parameters (typically three), see \cite{bayer2016pricing,fukasawa2011asymptotic,gatheral2014volatility}. One of the only potential drawbacks of such models in practice is the difficulty to price and hedge derivatives with them. Indeed, although some promising approaches have been recently introduced, see \cite{bennedsen2015hybrid}, due to the non-Markovian nature of the fractional Brownian motion, running efficient Monte-Carlo methods remains an intricate task in the rough volatility context, see \cite{neuenkirch2016order}.\\

\noindent However, it is shown in \cite{euch2016characteristic} that in the specific case of the so-called rough Heston model, instantaneous pricing of derivatives can be obtained. The rough Heston model of \cite{euch2016characteristic} is a natural extension\footnote{Actually there is no really standard definition for the rough Heston model and other versions can be considered, see \cite{guennoun2014asymptotic}.} to the rough framework of the classical Heston model of \cite{heston1993closed}. Indeed, the dynamic of the price $S$ on a probability space $(\Omega, {\cal{F}},\mathbb F ,\mathbb P)$ is defined as follows:
$$ dS_t = S_t \sqrt{V_t} dW_t $$
\begin{equation}
\label{roughHeston}
 V_t = V_0 +\frac{1}{\Gamma(\alpha)} \int_0^t (t-u)^{\alpha - 1} \lambda (\theta - V_u)  du  + \frac{1}{\Gamma(\alpha)} \int_0^t (t-u)^{\alpha - 1}  \nu \sqrt{V_u}  dB_u.
\end{equation}
Here the parameters $\lambda$, $\theta$, $V_0$, $S_0$ and $\nu$ are positive, $\alpha \in (1/2,1)$ and $W = \rho B + \sqrt{1-\rho^2} B^{\perp}$ with $(B,B^{\perp})$ a two-dimensional $\mathbb F$-Brownian motion and $\rho \in [-1,1]$. From \cite{euch2016characteristic}, the fractional stochastic differential equation \eqref{roughHeston} admits a unique weak solution and this solution has sample paths with H\"older regularity $\alpha - 1/2 - \varepsilon$ almost surely, for any $\varepsilon > 0$. Note also that in the case $\alpha=1$, we retrieve the classical Heston model. Surprisingly enough, it is proved in \cite{euch2016characteristic} that a semi-closed formula {\it \`a la Heston} also holds for the characteristic function of the log-price in the rough Heston model. This formula is very similar to that obtained in the classical Heston case, except that the classical time-derivative in the Riccati equation has to be replaced by a fractional derivative. Indeed, we have
$$ \mathbb{E}[\text{exp}\big(ia\log(S_T/S_0)\big)] = \text{exp}\big(g_1(a,t)  + V_0 g_2(a,t) \big),$$
where
$$ g_1(a,t) = \theta \lambda \int_0^t h(a,s) ds ,\quad g_2(a,t) = I^{1-\alpha}h(a,t),$$
and $h$ is the unique continuous solution of the following fractional Riccati equation: 
$$ D^\alpha h(a,s) = \frac{1}{2}(-a^2-ia)+  (i a \rho \nu- \lambda) h(a,s) + \frac{ \nu^2}{2 } h^2(a,s),~~I^{1-\alpha}h(a,0) = 0,$$
with $I^{1-\alpha}$ and $D^\alpha$ the fractional integral and derivative operators defined in Appendix \ref{fracCal}. When $\alpha=1$, this result does coincide with the classical Heston's result. Furthermore, efficient numerical pricing procedures for vanilla options can be easily designed from it, see \cite{euch2016characteristic}.\\

\noindent Thus, the relevance of the rough Heston model is twofold: it enjoys at the same time the nice modeling properties of rough volatility models and the computational advantages of the Heston framework.
However, the interest of having a pricing procedure is of course limited if it does not go along with a hedging strategy. Being able to build a hedging portfolio essentially means computing conditional expectations of the form $C_t=\mathbb E [f(S_T)|{\cal F}_t]$, where $f$ is a deterministic payoff function. In the classical Heston case, the Markovian structure of the model is very helpful to do it. In the rough case, this task is much more intricate since the underlying fractional Brownian motion is neither a Markov process nor a semi-martingale.\\

\noindent To tackle this issue, we first study the conditional laws in rough Heston models. We actually prove a very nice stability property. Indeed,
we show that conditional on ${\cal F}_t$, the law of the rough Heston model is still that of a rough Heston model, provided that the mean-reversion level $\theta$ is replaced by a time-dependent one. Hence we generalize our definition of the rough Heston model, allowing for the mean-reversion level to depend on time. Then using Hawkes processes as in \cite{euch2016characteristic}, we are able to compute the extended characteristic function of the log-price in generalized rough Heston models, that is
\begin{equation}\label{charfunc}
\mathbb{E}\big[\text{exp}\big(z\text{log}(S_t/S_0)\big)\big]
\end{equation}
for $z=a+ib$, with $b\in\mathbb{R}$ and $a$ in some subset of $\mathbb{R}$ to be defined later. From an explicit expression of \eqref{charfunc}, we can deduce a semi-closed formula for $C_t$, following for example the approach in \cite{carr1999option}.\\

\noindent Our most important result is the fact that we are able to identify the relevant state variables in rough Heston models, namely the underlying and the so-called forward variance curve: $(\mathbb E [V_{s+t}|{\cal F}_t])_{0 \leq s \leq T-t}$. 
Indeed, we show that $C_t$ can be written
$$ C_t = C\big(T-t,S_t,(\mathbb E [V_{s+t}|{\cal F}_t])_{s \geq 0}\big),$$ with $C()$ an explicit deterministic function. The above formula shows rigorously that the hedging instruments needed with rough models are the spot price and the forward variance curve, an idea already emphasized in \cite{bayer2016pricing}. Such result is also in the spirit of the 
approach developed in \cite{bergomi2005smile}. More precisely, we show that the dynamic of the option price satisfies
\begin{align*}
 dC_t &= \partial_S  C\big(T-t,S_t,(\mathbb E [V_{s+t}|{\cal F}_t])_{s \geq 0}\big) dS_t\\
 &+ \partial_V  C\big(T-t,S_t,(\mathbb E [V_{s+t}|{\cal F}_t])_{s \geq 0}\big).\big(d\mathbb E [V_{s+t}|{\cal F}_t])_{s \geq 0}\big),
\end{align*}
where $\partial_S C$ is the derivative of $C$ with respect to the underlying (the so-called delta) and $\partial_VC$ is the Fr\'echet derivative of $C$ according to the forward variance curve. From this expression, we readily obtain hedging strategies in 
terms of underlying and forward variance curve. Of course, in practice, one cannot really trade the whole forward variance curve. However, approximations can be built using liquid variance swaps or vanilla options.\\

\noindent Note also that using generalized rough Heston models enables us to perfectly fit the initial forward variance curve through the time varying mean-reversion parameter. Thus, one reproduces with great accuracy the dynamics of historical data, the whole implied volatility surface, including the at-the-money skew and the forward variance curve, and has access to instantaneous pricing and hedging methods.\\

\noindent The paper is organized as follows. In Section \ref{cond}, we investigate conditional laws of rough Heston models and introduce generalized rough Heston models with time-dependent mean-reversion level. Using Hawkes processes, we derive in Section \ref{characSection} the characteristic function of the log-price in generalized rough Heston models, emphasizing the role of the forward variance curve. We also discuss useful sufficient conditions for finite moments of the underlying price. Finally, we design our hedging strategies in Section \ref{hedgeSec}. Some proofs are relegated to Section \ref{proofs} and some technical results are given in an Appendix.

\section{Conditional laws of rough Heston models}\label{cond}
The goal of this paper is to understand how to price and hedge vanilla options with maturity $T>0$ and payoff $f(S_T)$ in the rough Heston framework \eqref{roughHeston}. Thus a first step is to characterize the law of the process $(S_t^{t_0},V_t^{t_0})_{t \geq 0} = (S_{t+{t_0}},V_{t+{t_0}})_{t \geq 0}$ conditional on ${\cal F}_{t_0}$, for a fixed ${t_0}>0$. Indeed, in order to derive the option price dynamic and to build hedging portfolios, one needs to be able to compute
$\mathbb E[f(S_T) | {\cal F}_t], \quad 0 \leq t \leq T$.\\ 

\noindent To state our result on conditional laws of rough Heston models, it is convenient to introduce a generalized version of Model \eqref{roughHeston}, allowing for time-varying mean-reversion level.  

\begin{definition}[Generalized rough Heston model] \label{GeneralRoughHeston}
On a filtered probability space  $(\Omega, {\cal{F}},\mathbb F ,\mathbb P)$, we define a generalized rough Heston model by
$$ dS_t = S_t \sqrt{V_t} dW_t $$
\begin{equation}
\label{hestonGeneral}
 V_t = V_0 +\frac{1}{\Gamma(\alpha)} \int_0^t (t-u)^{\alpha - 1} \lambda (\theta^0(u) - V_u)  du  + \frac{1}{\Gamma(\alpha)} \int_0^t (t-u)^{\alpha - 1}  \nu \sqrt{V_u}  dB_u.
\end{equation}
Here the parameters $\lambda$, $V_0$, $S_0$ and $\nu$ are positive, $\alpha \in (1/2,1)$ and $W = \rho B + \sqrt{1-\rho^2} B^{\perp}$ with $(B,B^{\perp})$ a two-dimensional $\mathbb F$-Brownian motion and $\rho \in [-1,1]$. Moreover, $\theta^0$ is a deterministic function, continuous on $\mathbb{R}_+^*$ satisfying \begin{equation} \label{condition1}
 \forall u>0 ; \quad  \theta^{0}(u)  \geq  - \frac{V_{0} }{\lambda \Gamma(1-\alpha)} u^{-\alpha},\end{equation}
and
\begin{equation} \label{condition2}\forall \varepsilon>0 \quad \exists K_\varepsilon>0; \quad \forall u \in (0,1] ; \quad \theta^{0}(u) \leq K_\varepsilon u^{-\frac{1}{2}-\varepsilon}.\end{equation}
\end{definition}
\noindent Note that under Conditions \eqref{condition1} and \eqref{condition2}, the fractional stochastic differential equation \eqref{hestonGeneral} admits a unique weak solution, see Theorem \ref{ConvergenceTheorem} and associated references.\\

\noindent We now give our result for the conditional laws of generalized rough Heston models (which Model \eqref{roughHeston} is a particular case of). Let $(S_t,V_t)_{t \geq 0}$ be defined by \eqref{hestonGeneral}. We have the following theorem, proved in Section \ref{proofLaw}.

\begin{theorem} \label{condLaw}
The law of the process $(S^{t_0}_t,V^{t_0}_t)_{t \geq 0} $ is that of a generalized rough Heston model with the following dynamic:
$$ dS_t^{t_0} = S_t^{t_0} \sqrt{V_t^{t_0}} dW_t^{t_0}, \quad S_0^{t_0} = S_{t_0}$$
$$ V_t^{t_0} = V_{t_0} + \frac{1}{\Gamma(\alpha)} \int_0^t (t-u)^{\alpha - 1} \lambda (\theta^{t_0}(u) - V_u^{t_0})  du  + \frac{1}{\Gamma(\alpha)} \int_0^t (t-u)^{\alpha - 1}  \nu \sqrt{V_u^{t_0}}  dB_u^{t_0},$$
with $(W^{t_0}_t,B^{t_0}_t)_{t \geq 0} = (W_{{t_0}+t}-W_{t_0},B_{{t_0}+t}-B_{t_0})_{t \geq 0}$ a two-dimensional Brownian motion with correlation $\rho$, independent of ${\cal F}_{t_0}$ and
$$ \theta^{t_0}(u) = \theta^0(t_0+u) + \frac{\alpha}{\lambda \Gamma(1-\alpha)} \int_0^{t_0} ({t_0}-v+u)^{-1-\alpha}  (V_v - V_{t_0})dv  +  \frac{(u+{t_0})^{-\alpha}}{\lambda \Gamma(1-\alpha)} (V_0 - V_{t_0}),$$
which is an ${\cal F}_{t_0}$-measurable function continuous on $\mathbb{R}_+^*$ such that Conditions \eqref{condition1} and \eqref{condition2} (where the index $0$ should be replaced by $t_0$) are satisfied.
\end{theorem}

\noindent Hence the class of generalized rough Heston models is stable with respect to conditioning. The conditional law of a rough Heston model is still that of a rough Heston model. The only difference is a modification in the mean-reversion level function. In particular, when considering the usual rough Heston model \eqref{roughHeston}, the constant parameter $\theta$ becomes an $\mathcal{F}_{t_0}$ measurable function when taking conditional law at time $t_0$. This result will be crucial to derive hedging strategies in the rough Heston framework, and more generally to understand the state variables associated to rough Heston type dynamics.

\section{Characteristic function of generalized rough Heston models}\label{characSection}

The goal of this section is to derive the extended characteristic functions for the log-price in the rough Heston model \eqref{hestonGeneral}. This together with Theorem \ref{condLaw} will enable us to derive conditional characteristic functions, leading to hedging strategies. The first step to achieve this goal is to build a suitable sequence of processes converging to the generalized rough Heston model of Definition \ref{GeneralRoughHeston}. Then we will be able to do computations on these processes (notably deriving characteristic functions), and pass them to the limit to obtain results for generalized rough Heston models.\\

\subsection{Generalized rough Heston models as limit of nearly unstable Hawkes processes}

\noindent In \cite{euch2016characteristic}, a microscopic price model, based on two-dimensional Hawkes processes, is built so that it converges on the long run after suitable rescaling to a rough Heston log-price (with constant mean-reversion). Then, characteristic functions are obtained from this result. Such method could easily be extended to obtain a generalized rough Heston model in the limit. However, it would only enable us to compute \eqref{charfunc} with $a=0$. This is not enough so that classical Fourier inversion methods such as that in \cite{carr1999option} can be rigorously applied to compute prices and hedging portfolios.\\ 

\noindent Thus we use another approach in this section, quite similar to that of \cite{jaisson2016rough}. We consider a sequence of one-dimensional Hawkes processes
$(N^T_t)_{t \geq 0}$, indexed by $T>0$ going to infinity, with intensity given by
$$ \lambda_t^T = \mu_T+ \int_0^t a_T \varphi(t-s) dN_s^T,$$
where $\mu_T$ and $a_T$ are positive constants with $a_T<1$ and $\varphi : \mathbb R_+^* \rightarrow \mathbb R_+$ is integrable such that $\int_0^\infty \varphi = 1$. In \cite{jaisson2016rough}, it is shown that provided
\begin{equation}\label{Hcond1}
x^{\alpha} \int_x^\infty \varphi(s) ds \underset{x \rightarrow \infty}{\longrightarrow} \frac{1}{\Gamma(1-\alpha)}, \quad  \alpha \in (1/2,1),
\end{equation}
and 
\begin{equation*}
T^{\alpha}(1-a_T) \underset{T \rightarrow \infty}{\longrightarrow} \lambda,\quad T^{1-\alpha}\mu_T \underset{T \rightarrow \infty}{\longrightarrow} \lambda/\nu^2,
\end{equation*}
for some positive constants $\lambda$ and $\nu$, a suitably rescaled version of the intensity process $\lambda_t^T$ asymptotically behaves as the variance process of a rough Heston model with constant mean-reversion parameter such as 
\eqref{roughHeston} and with initial variance equal to zero. To obtain a time-dependent mean-reversion level and a non-zero starting value in the limit, we are inspired by an idea in \cite{euch2016characteristic}, where it is shown that a time-dependent $\mu_T$ is a way to modify some parameters in the limit. More precisely, we consider the following assumption, where $f^{\alpha,1}$ denotes the 
Mittag-Leffler density function defined in Appendix \ref{mittag}.
\begin{assumption}\label{assump1Hawkes} There exist $\lambda, \nu>0$, $\alpha \in (1/2,1)$ and $V_0>0$ such that for $T>1/\lambda^{-1/\alpha}$ and $t\geq 0$,
$$ \lambda_t^T = \mu_T \zeta^T(t)+ \int_0^t \varphi^T(t-s) dN_s^T,$$
where 
$$ a_T = 1 - \lambda T^{-\alpha}, \quad \mu_T = (\lambda/\nu^2 )T^{\alpha-1}, \quad\varphi^T = a_T \varphi,$$ with $\varphi=f^{\alpha,1}$ and
$$ \zeta^T(t) = V_0 \big( \frac{1}{1-a_T} (1 - \int_0^t \varphi^T(t-s) ds ) - \int_0^t   \varphi^T(t-u)du\big)  + \int_0^t   \varphi^T(t-u) \theta^0(u/T) du,$$ 
where $\theta^0()$ satisfies the assumptions of Definition \ref{GeneralRoughHeston}.
\end{assumption}
\noindent Note that we are working in the so-called nearly unstable case for Hawkes processes since the $L^1$ norm of the kernel $\varphi^T$ converges to one. Furthermore remark that $\eqref{Hcond1}$ is satisfied, see Appendix \ref{mittag}.

\begin{rem}
\label{mu}
Remark that $\zeta^T$ can also be written as follows
$$ \zeta^T(t) =  \int_0^t   \varphi^T(t-u) \theta^0(u/T) du + V_0  \big(\frac{T^\alpha}{\lambda}  \int_t^\infty \varphi(s) ds  + \lambda T^{-\alpha} \int_0^t \varphi(s) ds \big).$$
Therefore using that $I^{1-\alpha}\varphi^T(t) = \int_t^{\infty} \varphi^T$, see Appendix \ref{mittag}, together with Condition \eqref{condition1} we get
\begin{align*}
\zeta^T(t) &\geq - \frac{V_0}{\lambda \Gamma(1-\alpha)}T^\alpha \int_0^t   \varphi^T(t-u) u^{-\alpha} du + V_0    \big(\frac{T^\alpha}{\lambda}  \int_t^\infty \varphi(s) ds  + \lambda T^{-\alpha} \int_0^t \varphi(s) ds \big)\\
&= - \frac{V_0}{\lambda}T^\alpha \int_t^\infty   \varphi^T(s) ds + V_0    \big(\frac{T^\alpha}{\lambda}  \int_t^\infty \varphi(s) ds  + \lambda T^{-\alpha} \int_0^t \varphi(s) ds \big)\\
&=  V_0\mu_T ( \int_t^\infty   \varphi(s) ds + \lambda T^{-\alpha} \int_0^t \varphi(s) ds \big).
\end{align*}
\noindent This shows that $\zeta^T$ is a positive function and thus that the intensity process $\lambda_t^T$ is well-defined.
\end{rem}

\noindent 
We define $M_t^T=N_{t}^T-\int_0^t\lambda_s^T ds$ and
$$X_t^T= \nu^2\frac{1-a_T}{T^\alpha\lambda} N_{tT}^T,~\Lambda_t^T=  \nu^2\frac{1-a_T}{T^\alpha \lambda} \int_0^{tT} \lambda_{s}^T ds,~Z_t^T = \nu\sqrt{\frac{1-a_T}{T^\alpha \lambda}} M_{tT}^T.$$
Conditions \eqref{condition1} and \eqref{condition2} on the function $\theta^0$ allow us to adapt the proofs in \cite{euch2016characteristic,jaisson2016rough} in a straightforward way to obtain the following result.

\begin{theorem}
\label{ConvergenceTheorem}
Let $t_0>0$. As $T \rightarrow \infty$, under Assumption \ref{assump1Hawkes}, the process $\big{(} \Lambda_t^T, X_t^T, Z_t^T \big{)}_{t \in [0,t_0]}$ converges in law for the Skorokhod topology to $(\Lambda, X, Z)$, where
\begin{itemize}
\item $\displaystyle \Lambda_t = X_t = \int_0^t V_s ds.$ 
\item $\displaystyle Z_t = \int_0^t \sqrt{V_s} dB_s$ which is a continuous martingale.
\item $V$ is the unique weak solution of the rough stochastic differential equation 
$$ V_t = V_0 +  \frac{1}{\Gamma(\alpha)} \int_0^t (t-s)^{\alpha - 1} \lambda(\theta^0(s)-V_s) ds + \frac{\nu}{\Gamma(\alpha)} \int_0^t (t-s)^{\alpha - 1} \sqrt{V_s} dB_s,$$
\end{itemize}
where $B$ is a Brownian motion. Furthermore, the process $V$ is non-negative and has H\"older regularity $\alpha - 1/2 - \varepsilon$ for any $\varepsilon >0$.
\end{theorem}
\noindent Theorem \ref{ConvergenceTheorem} will be one of the key results to obtain the extended characteristic function of the log-price in generalized rough Heston models.

\subsection{Conditions for finite moments in generalized rough Heston models}\label{condFiniteMoment}

\noindent Recall that we aim at computing \eqref{charfunc} with $\Re(z)\neq 0$. A preliminary step towards this is to derive sufficient conditions for the finiteness of the moments of $S_t$ and 
$\text{exp}\big(\int_0^tV_s ds\big)$ in generalized rough Heston models. To obtain such result, we use Theorem \ref{ConvergenceTheorem}. Let $a\in\mathbb{R}$. First, note that
$$ (S_t)^a = (S_0)^a \text{exp}\big(a \rho \int_0^t \sqrt{V_s} dB_s - \frac{a}{2} \int_0^t V_s ds + a \sqrt{1-\rho^2} \int_0^t \sqrt{V_s} dB_s^{\perp}\big).$$
Consequently, we have
$$ \mathbb E [(S_t)^a] =(S_0)^a \mathbb E \big[ \text{exp}(a \rho \int_0^t \sqrt{V_s} dB_s + \frac{1}{2}(-a + a^2 (1-\rho^2)) \int_0^t V_s ds)\big].$$
Now define
$$ M_t = \text{exp}\big( a \rho \int_0^t \sqrt{V_s} dB_s- \frac{a^2 \rho^2}{2} \int_0^t V_s ds \big ).$$
The process $M_t$ is a positive local martingale and actually, by Proposition \ref{martingale} in Appendix, a true martingale. Define the corresponding probability measure $\mathbb Q$:
$$ \frac{d \mathbb Q}{d \mathbb P}\Big |_{{\cal F}_t} = M_t. $$
By Girsanov theorem, under $\mathbb Q$, 
$$ B^{\mathbb Q}_t = B_t - a \rho \int_0^t \sqrt{V_s} ds  $$
is a $\mathbb F$-Brownian motion. Consequently, under $\mathbb Q$, $V$ defined in \eqref{hestonGeneral} is still the variance process of a generalized rough Heston model, but with different parameters:
\begin{equation}\label{underQ} 
V_t = V_0 +\frac{1}{\Gamma(\alpha)} \int_0^t (t-u)^{\alpha - 1} \tilde{\lambda}( \tilde{\theta^0}(u) - V_u)  du  + \frac{1}{\Gamma(\alpha)} \int_0^t (t-u)^{\alpha - 1}  \nu \sqrt{V_u}  dB_u^{\mathbb Q}, 
\end{equation} 
where
$$ \tilde{\lambda} = \lambda - \rho \nu a, \quad \tilde{\theta^0}(t) = \frac{\lambda \theta^0(t)}{\lambda - \rho \nu a},$$
provided that $ \lambda - \rho \nu a > 0$.
Hence we obtain 
\begin{equation}\label{linkCond12}
\mathbb E [(S_t)^a] =(S_0)^a \mathbb E_{\mathbb Q} [ \text{exp}( \frac{1}{2}(-a + a^2 ) \int_0^t V_s ds) ].
\end{equation}
Therefore, a sufficient condition on $a$ for
\begin{equation}\label{condMoment1}
\mathbb E_{\mathbb Q}[ \text{exp}( \frac{1}{2}(-a + a^2 ) \int_0^t V_s ds) ]<\infty
\end{equation}
 will readily imply a sufficient condition for the finiteness of $\mathbb E [(S_t)^a]$.\\

\noindent We now explain how to derive such condition. Recall that from Theorem \ref{ConvergenceTheorem}, $ \nu^2 T^{-2\alpha}N^T_{tT} $
converges in law to $\int_0^t V_s ds$. Thus we look first for a condition on $a \in \mathbb R$ for which
\begin{equation} \label{condMoment3} 
\mathbb E [\text{exp}(a\nu^2  T^{-2\alpha}N^T_{tT})] < \infty ,
\end{equation}
for large enough $T>0$ and fixed $t>0$. This is done using a population interpretation of Hawkes processes, see Appendix \ref{branching}. It leads us to a sufficient condition on $a \in \mathbb R$ for \eqref{condMoment1}. Furthermore, we are able to compute explicitly the expectation in \eqref{condMoment3}, see Appendix \ref{branching}. Thus we can pass to the limit as $T$ goes to infinity and then obtain an explicit expression for the expectation in \eqref{condMoment1}. More precisely, we have the following result whose proof is given in Section \ref{proofMoment1}, where $a_0(t)$ is defined for $t>0$ by
$$a_0(t) =  \frac{1}{2 \nu^2} (\lambda + \frac{\alpha t^{-\alpha}}{\Gamma(1-\alpha)})^2.$$

\begin{theorem}\label{moment1}
Let $V$ be the variance process of the generalized rough Heston model \eqref{hestonGeneral}. For any $t>0$ and 
$ a < a_0(t)$,
$$ \mathbb{E}\big[\emph{exp}(a \int_0^t V_s ds)\big] < \infty $$
and 
$$ \mathbb{E}\big[\emph{exp}(a \int_0^t V_s ds)\big] = \emph{exp}\big(\int_0^t g(a,t-s) (\lambda \theta^0(s) + \frac{V_0 s^{-\alpha}}{\Gamma(1-\alpha)}) ds\big),$$
where $g(a,.)$ is the unique continuous solution of the following fractional Riccati equation:
$$ D^\alpha g(a,s) = a - \lambda g(a,s) + \frac{\nu^2}{2}g(a,s)^2,\quad s\leq t, \quad I^{1-\alpha}g(a,0)= 0.$$
For any $0 \leq s\leq t$, this function satisfies
$$ g(a,s) \leq \frac{c}{\nu^2}  \big( \frac{\alpha s^{-\alpha}}{\Gamma(1-\alpha)}+\nu\sqrt{a_0(s)-a} \big)$$
for some constant $c>0$. Furthermore, for fixed $0\leq s\leq t$,
 $a \rightarrow g(a,s) $ is non-decreasing and $s \rightarrow g(a,s)$ is non-increasing on $[0,t]$ if $a<0$ and non-decreasing if $a>0$.
\end{theorem}

\noindent Let $S_t$ denote the price in the generalized rough Heston model of Definition \ref{GeneralRoughHeston}. Using \eqref{linkCond12}, we obtain the following corollary on the moments of $S_t$.

\begin{corollary}\label{moment2} Let $t>0$.
Assume $$ \lambda - \rho \nu a>0, \quad a_-(t) < a < a_+(t), $$
where $$ a_-(t) = \frac{\nu^2 - 2 \rho \nu X(t) + \sqrt{\Delta(t)}}{2 \nu^2(1-\rho^2)},\quad a_+(t) = \frac{\nu^2 - 2 \rho \nu X(t) - \sqrt{\Delta(t)}}{2 \nu^2(1-\rho^2)},$$ 
with $$ X(t) = \lambda+\frac{\alpha t^{-\alpha}}{ \Gamma(1-\alpha)}, \quad \Delta(t) = 4 \nu^2 X(t)^2 + \nu^4 - 4 \rho \nu^3 X(t).$$
Then we have $$ \mathbb E [(S_t)^a] < \infty.$$ Furthermore, 
$$\mathbb E [(S_t)^a]  = (S_0)^a \emph{exp}\big(\int_0^t h(a,t-s) (\lambda \theta^0(s) + \frac{V_0 s^{-\alpha}}{\Gamma(1-\alpha)}) ds\big) ,$$
where $h(a,.)$ is the unique continuous solution of the following fractional Riccati equation:
$$D^\alpha h(a,s) = \frac{a^2-a}{2} - (\lambda - \rho \nu a) h(a,s) + \frac{\nu^2}{2}h(a,s)^2,\quad s\leq t,\quad I^{1-\alpha}h(a,0)= 0.$$
\end{corollary}

\begin{remark} Note that if we formally take $\alpha = 1$ in Corollary \ref{moment2}, our model coincides with the classical Heston model. In that case $X(t) = \lambda$ and therefore $a_-$ and $a_+$ do not depend on $t$. Moreover the set of $a \in \mathbb R$ such that 
$$\lambda - \rho \nu a>0, \quad  a_- \leq a \leq a_+,$$
exactly corresponds to that of $a\in \mathbb R$ for which
$$ \forall t \geq 0, \quad \mathbb E [(S_t)^a] < \infty,$$
see \cite{andersen2007moment} for further details on moment explosions for the classical Heston model.
\end{remark}

\noindent\textsc{Proof of Corollary \ref{moment2}}:\\

\noindent Recall that from \eqref{linkCond12},
$$ \mathbb E [(S_t)^a] =(S_0)^a \mathbb E_{\mathbb Q} [ \text{exp}( \frac{1}{2}(-a + a^2 ) \int_0^t V_s ds) ] .$$
From Theorem \ref{moment1} and the fact that under $\mathbb Q$, $V$ follows \eqref{underQ}, this quantity is finite if  $\lambda - \rho \nu a > 0 $ and 
$$ \frac{1}{2}(-a + a^2 ) < \tilde{a}_0(t)= \frac{1}{2 \nu^2} (\tilde{\lambda}  + \frac{\alpha t^{-\alpha}}{\Gamma(1-\alpha)})^2 = \frac{1}{2 \nu^2} (\lambda -\rho \nu a  + \frac{\alpha t^{-\alpha}}{\Gamma(1-\alpha)})^2.$$
This is equivalent to
$$ a^2 \nu^2 (1-\rho^2) +a (-\nu^2 + 2 X(t) \rho \nu) - X(t)^2 < 0.$$
The conditions on $a \in \mathbb R$ stated in Corollary \ref{moment2} follow. Finally, the expression of $\mathbb E [(S_t)^a] $ is easily obtained using \eqref{linkCond12} together with Theorem \ref{moment1}.
\qed

\subsection{Characteristic functions of generalized rough Heston models} \label{CharacSection}
We are now ready to derive the characteristic functions of generalized rough Heston models. Let $t>0$.
We want to compute $$ R(z,t) = \mathbb E\big[\text{exp}\big(z \log(S_t/S_0)\big)\big],  $$
where $z \in \mathbb C$ satisfies
\begin{equation}\label{condZ}
z = a + i b, \quad a, b \in \mathbb R, \quad \lambda - \rho \nu a>0, \quad a_-(t) < a < a_+(t),
\end{equation}
where $a_-(t)$ and $a_+(t)$ are defined in Corollary \ref{moment2}. Recall that from Corollary \ref{moment2}, \eqref{condZ} implies that $\text{exp}\big(z \log(S_t/S_0)\big)$ is integrable and therefore $R(z,t)$ is well-defined.\\

\noindent Using the same computations as in the preceding sections, we get
\begin{equation} \label{linkCharac12} 
R(z,t) = \mathbb E_{\mathbb Q}\big[\text{exp} \big( ib \rho \int_0^t \sqrt{V_s} dB_s^{\mathbb Q} +\frac{1}{2} ( \rho^2 b^2 + {z^2}  - z) \int_0^t V_s ds   \big)\big].
\end{equation}
As already seen, under $\mathbb Q$, $V$ still follows the variance process of a generalized rough Heston model driven by the Brownian motion $B^{\mathbb Q}$, see \eqref{underQ}. Thus, we need to study
$$ G(z,x,t) = \mathbb E \big[\text{exp}\big( ix \int_0^t \sqrt{V_s} dB_s + z \int_0^t V_s ds \big)\big], 
$$
with $x \in \mathbb R$, $z \in \mathbb C$ such that $\Re (z) < a_0(t)$, ($a_0(t)$ is defined in Theorem \ref{moment1}), and $V$ is the variance process of a generalized rough Heston model. To do so, we use again Theorem \ref{ConvergenceTheorem}. Indeed
$(\nu^2T^{-2\alpha}N_{tT}^T, \nu T^{-\alpha}M_{tT}^T)$ converges in law as $T$ goes to infinity to $(\int_0^t V_s ds , \int_0^t \sqrt{V_s} dB_s)$. Computing 
$$ \mathbb E [\text{exp}\big(ix \nu T^{-\alpha}M_{tT}^T + z \nu^2T^{-2\alpha}N_{tT}^T  \big)] $$
and passing to the limit, we obtain the following result whose proof is given in Section \ref{proofMoment3}.

\begin{theorem}\label{moment3}
Let $V$ be the variance process of the generalized rough Heston model \eqref{hestonGeneral}. For any $t>0$, $b \in \mathbb R$ and $z \in \mathbb C$ such that  $ \Re(z) < a_0(t)$,
$$ G(z,x,t) = \emph{exp}\big(\int_0^t \xi(z,x,t-s) (\lambda \theta^0(s) + \frac{V_0 s^{-\alpha}}{\Gamma(1-\alpha)}) ds\big),$$
where $\xi(z,x,.)$ is the unique continuous solution of the following fractional Riccati equation:
$$ D^{\alpha} \xi(z,x,s) = z - \frac{x^2}{2} + (ix \nu - \lambda) \xi(z,x,s) +\frac{\nu^2}{2}\xi(z,x,s)^2,\quad s\leq t, \quad I^{1-\alpha}\xi(z,x,0) = 0.$$
\end{theorem}

\noindent The following corollary is readily obtained from Theorem \ref{moment3} together with \eqref{linkCharac12}.
\begin{corollary}\label{moment4} Let $t>0$ and $z \in \mathbb C$ satisfying \eqref{condZ}. We have 
$$ R(z,t)  = \emph{exp}\big(\int_0^t h(z,t-s) (\lambda \theta^0(s) + \frac{V_0 s^{-\alpha}}{\Gamma(1-\alpha)}) ds \big), $$
where $h(z,.)$ is the unique continuous solution of the following fractional Riccati equation:
$$ D^{\alpha}h(z,s) =\frac{1}{2}( {z^2}  - z) + (z \rho \nu - \lambda ) h(z,s) + \frac{\nu^2}{2} h(z,s)^2,\quad s\leq t, \quad I^{1-\alpha}h(z,0) = 0.$$
\end{corollary}

\subsection{Connection with the forward variance curve}

We now show how the characteristic function given in Corollary \ref{moment4} can be written as a functional of the forward variance curve $(\mathbb E [V_t])_{t \geq 0}$. This property will be crucial in the next section when computing hedging portfolios. 
We first remark that the time-dependent parameter $\theta^0$ can be directly linked to the forward variance curve through the following result. 

\begin{prop}\label{linkThetaForward} 
Let $V$ be the variance process of the generalized rough Heston model \eqref{hestonGeneral}. For any $t\geq 0$, we have
\begin{equation} \label{expectationV}
\mathbb E[V_t] = V_0\big(1-F^{\alpha,\lambda}(t)\big) + \int_0^t f^{\alpha,\lambda}(t-s) \theta^0(s) ds, 
\end{equation}
where $F^{\alpha,\lambda}$ and $f^{\alpha,\lambda}$ are defined in Appendix \ref{mittag}. Furthermore, $\theta^0$ can be written as a functional of the forward variance curve as follows:
\begin{equation} \label{thetaForward} 
\lambda \theta^0(t) +  V_0 \frac{t^{-\alpha}}{\Gamma(1-\alpha)}  = D^{\alpha} \mathbb E [V_t]  +  \lambda \mathbb E [V_t], \quad t>0.
\end{equation}
\end{prop}

\noindent\textsc{Proof of Proposition \ref{linkThetaForward}:}\\
 
\noindent In the same way as in \cite{jaisson2016rough}, we can show that for any $t \geq 0$, 
$$\mathbb E [\int_0^t V_s ds] < \infty.$$
So we have that $t \rightarrow \mathbb E [V_t]$ is locally integrable. Moreover $f^{\alpha,\lambda}$ is square-integrable, see Appendix \ref{mittag}. Thus we obtain that for any $t \geq 0$, 
$$ \int_0^t  f^{\alpha,\lambda}(t-s)^2 \mathbb E [ V_s]  ds < \infty.$$
Therefore,
$$\mathbb E [ \int_0^t  f^{\alpha,\lambda}(t-s)  \sqrt{V_s}  dB_s ]= 0.$$
Writing the dynamic of $V$ under the following form as in \cite{jaisson2016rough}:
\begin{equation}\label{dynV}
 V_t= V_0\big(1-F^{\alpha,\lambda}(t)\big) + \int_0^t f^{\alpha,\lambda}(t-s) \theta^0(s) ds+ \frac{\nu}{\lambda}\int_0^t f^{\alpha,\lambda}(t-s) \sqrt{V_s} dB_s,
\end{equation}
we deduce \eqref{expectationV}. Now using Fubini theorem and noting that $ I^{1-\alpha} f^{\alpha,\lambda} = \lambda (1-F^{\alpha,\lambda}) $, see Appendix \ref{mittag}, we get that for any $t \geq 0$, 
$$I^{1-\alpha} \mathbb E[V_t] =  V_0 \frac{t^{1-\alpha}}{(1-\alpha)\Gamma(1-\alpha)} + \int_0^t \lambda \big(1- F^{\alpha,\lambda}(t-s)\big) (\theta^0(s)-V_0) ds.$$
Using Fubini again, this can be rewritten
$$I^{1-\alpha} \mathbb E[V_t] =  V_0 \frac{t^{1-\alpha}}{(1-\alpha)\Gamma(1-\alpha)} + \int_0^t \lambda (\theta^0(s)-V_0) ds - \lambda \int_0^t  \int_0^s f^{\alpha,\lambda}(s-u) (\theta^0(u)-V_0) du ds.$$
Then from \eqref{expectationV} we derive
$$I^{1-\alpha} \mathbb E[V_t] =  V_0 \frac{t^{1-\alpha}}{(1-\alpha)\Gamma(1-\alpha)} + \int_0^t \lambda (\theta^0(s)-V_0) ds - \lambda \int_0^t  (\mathbb E[V_s] - V_0) ds.$$ 
We finally obtain \eqref{thetaForward} by differentiating this last equality.\\
\qed

\begin{remark} Assume that the forward variance curve $t \rightarrow \mathbb E[V_t]$ is observed on the market through the implied volatility surface or liquid variance swaps, 
and that this curve admits a fractional derivative of order $\alpha$. Then the mean-reversion function $\theta^0$ can be chosen so that the model is consistent with this market forward variance curve by taking
$$ \lambda \theta^0(t) = D^{\alpha} (\mathbb E[V_.] - V_0)(t) + \lambda \mathbb E [V_t] .$$
\end{remark}

\noindent From Corollary \ref{moment4} together with Proposition \ref{linkThetaForward}, 
we can eventually write the characteristic function of the log-price as a functional of the forward variance curve. 
Thus, it indicates that the forward variance curve is a relevant state variable in generalized rough Heston models. Such type of phenomena also appears in the class of models developed in \cite{bergomi2005smile}. More precisely, we have the following corollary.

\begin{corollary} \label{moment5}
Let $t>0$ and $z \in \mathbb C$ satisfying \eqref{condZ}. We have 
$$ R(z,t)= \emph{exp}\big(\int_0^t \chi(z,t-s) \mathbb E[V_s] ds \big),$$
where 
$$ \chi(z,t) = \frac{1}{2}( {z^2}  - z) + z \rho \nu h(z,t) + \frac{\nu^2}{2} h(z,t)^2,$$
with $h(z,.)$ the unique continuous solution of the fractional Riccati equation given in Corollary \ref{moment4}.
\end{corollary}
\noindent Thus, characteristic functions, and therefore conditional characteristic functions of the log-price can be written in term of the forward variance curve.
This shows that this object plays the role of state variable in this infinite dimensional fractional setting. Actually, this result could probably be understood in a more general framework of affine processes, see \cite{abi2017affine,cuchiero2017affine}.\\
  
\noindent\textsc{Proof of Corollary \ref{moment5}}:\\
 
\noindent By Lemma \ref{EDPFracLin} in Appendix, for any $0\leq s\leq t$,
\begin{equation}\label{hchi}
h(z,s) = \int_0^s \frac{1}{\lambda} f^{\alpha,\lambda}(s-u) \chi(z,u) du. 
\end{equation}
Moreover, from \eqref{expectationV} together with the fact that $ I^{1-\alpha} f^{\alpha,\lambda} = \lambda (1-F^{\alpha,\lambda}) $, see Appendix \ref{mittag}, we have
$$ \mathbb E [V_s] = \int_0^s \frac{1}{\lambda} f^{\alpha,\lambda}(s-u) (\lambda \theta^0(u) + V_0 \frac{ u^{-\alpha}}{\Gamma(1-\alpha)} ) du .$$
Then, using Fubini theorem, we obtain 
$$ \int_0^t \mathbb \chi(z,t-s) \mathbb E [V_s] ds =  \int_0^t \big(\int_0^{t-s}  \frac{1}{\lambda} f^{\alpha,\lambda}(t-s-u) \chi(z,u) du\big) (\lambda \theta^0(s) + V_0 \frac{ s^{-\alpha}}{\Gamma(1-\alpha)} ) ds$$
and therefore
$$ \int_0^t \mathbb \chi(z,t-s) \mathbb E [V_s] ds =  \int_0^t h(z,t-s) (\lambda \theta^0(s) + V_0 \frac{ s^{-\alpha}}{\Gamma(1-\alpha)} ) ds.$$
The result follows from Corollary \ref{moment4}.
\qed

\section{Hedging under generalized rough Heston models} \label{hedgeSec}
We consider a generalized rough Heston model with the additional assumption that $\rho \leq 0 $. We show in this section how to compute explicitly hedging portfolios for vanilla options in such model. We treat here the case of a European
call option with maturity $T>0$ and strike $K > 0$. Nevertheless, the approach can be easily extended to other vanilla payoffs.\\

\noindent It is easy to see that we can find $a>1$ such that the conditions of Corollary \ref{moment2} are satisfied for any $t \geq 0$. Therefore, for any $t\geq 0$, 
$$\mathbb E[(S_t)^a]<\infty.$$
We define the call option price process
$$ C_t = \mathbb E [(S_T-K)_+|{\cal F}_t], \quad 0 \leq t \leq T. $$
We write
$$X_t = \log(S_t), \quad t \geq 0$$ 
and $$ g(x) =   e^{-ax} (e^x-K)_+, \quad x \in \mathbb R. $$
We have $g \in\mathbb L^1(\mathbb R) \cap \mathbb L^2(\mathbb R)$ and therefore
$$ g(x) = \frac{1}{2 \pi} \int_{b \in \mathbb R} \hat{g}(-b) e^{ibx} db,$$
where $\hat{g} \in\mathbb L^1(\mathbb R) \cap  \mathbb L^2(\mathbb R)$ is the Fourier transform of $g$. Note that we are able to compute explicitly $\hat{g}$:
$$ \hat{g}(b) = \frac{e^{(1-a+ib)\text{log}(K)}}{(ib-a)(ib-a+1)}, \quad b \in \mathbb R. $$  
We then deduce by Fubini theorem that
\begin{equation} \label{parseval} 
C_t = \mathbb{E}[g(X_T)e^{a X_T}|{\cal F}_t] =  \frac{1}{2 \pi} \int_{b \in \mathbb R} \hat{g}(-b) P_t^T(a+ib) db,
\end{equation}
where 
$$ P^T_t(a+ib) = \mathbb E [\text{exp}\big( (a+ib) X_T \big) | {\cal F}_t] .$$

\noindent Using the fact that conditional on ${\cal F}_t$, $S$ still follows a generalized rough Heston dynamic together with Corollary \ref{moment5}, we obtain
$$ \mathbb E[\text{exp}\big((a+ib)\log(S_T/S_t)|{\cal F}_t\big)] = \text{exp}\big(\int_0^{T-t} \chi(a+ib,T-t-s) \mathbb E[V_{s+t}|{\cal F}_t] ds \big),$$ where $\chi$ is  defined in Corollary \ref{moment5}. 
Thus,
\begin{equation}\label{characProcess}
P^T_t(a+ib) = \text{exp}\big((a+ib) \log(S_t)+ \int_0^{T-t} \chi(a+ib,T-t-s) \mathbb E[V_{s+t}|{\cal F}_t] ds  \big).
\end{equation}
Hence, from \eqref{characProcess}, we deduce that $P^T_t(a+ib)$ is a deterministic functional of the underlying spot price $S_t$ and the forward variance curve until maturity $T$: $\mathbb E [V_{t+u}|{\cal F}_t],\quad  0 \leq u \leq T-t $.\\ 

\noindent Let 
$$ {\cal V}_{\alpha,\lambda} = \{ \xi : \mathbb R_+ \rightarrow \mathbb R_+ , \quad \xi(t) = \int_0^t \frac{s^{-\alpha}}{\lambda \Gamma(1-\alpha)} f^{\alpha,\lambda}(t-s)  \theta_\xi(s) ds, \quad \theta_\xi \text{ is continuous on } \mathbb R_+   \}.$$
The space ${\cal V}_{\alpha,\lambda}$ is a metric space containing 
$$ {\cal V}_{\alpha,\lambda}^+ = \{ \xi \in {\cal V}_{\alpha,\lambda}, \quad \theta_\xi >0  \text{ and for any } t > 0,\quad \theta_\xi (t) = \xi(0) + t^\alpha \lambda \Gamma(1-\alpha) \theta_\xi^0(t), \quad  \theta_\xi^0 \text{ satisfies \eqref{condition2}} \},$$
which is the set of all possible forward variance curves produced by generalized rough Heston models. Note that from the same computations as for Proposition \ref{linkThetaForward}, 
we get the uniqueness of the function $\theta_\xi$ for each $\xi \in {\cal V}_{\alpha,\lambda}$ since we have
\begin{equation*}\theta_\xi(t)   = \big(D^\alpha \xi(t) + \lambda \xi(t)\big) \Gamma(1-\alpha) t^\alpha, \quad t>0.\end{equation*}
We equip $ {\cal V}_{\alpha,\lambda}$ with the following complete metric:
$$d_{\alpha,\lambda}(\xi ,\zeta) =  \| |\theta_\xi - \theta_\zeta| \wedge 1 \|_\infty.$$
From \eqref{parseval} and \eqref{characProcess}, we 
get that the spot price and the forward variance curve are the relevant state variables for the call price process. Indeed, there exists a deterministic functional $C : \mathbb R_+ \times \mathbb R_+^* \times {\cal V}_{\alpha,\lambda} \rightarrow \mathbb R$ such that
$$ C_t = C\big(T-t, S_t, (\mathbb E[V_{s+t}|{\cal F}_t])_{ s \geq 0} \big), \quad t \in [0,T], $$
where for any $t \geq 0$, $S \in \mathbb R_+$ and $\xi \in {\cal V}_{\alpha,\lambda}$
\begin{equation} \label{functional}
C\big(t,S,\xi\big) =  \frac{1}{2 \pi} \int_{b \in \mathbb R} \hat{g}(-b) L(a+ib,t,S,\xi) db,
\end{equation}
with
$$ L(a+ib,t,S,\xi) = \text{exp}\big((a+ib) \log(S)+ \int_0^{t} \chi(a+ib,t-s) \xi(s)ds \big).$$

\noindent In the following proposition, proved in Section \ref{proofregu}, we give some useful regularity properties of the functional $C$.
\begin{prop} \label{regularity} Let $\xi \in {\cal V}^+_{\alpha,\lambda}$, $S>0$, $t>0$ and assume $|\rho|<1$. The function $C(t,.,\xi)$ defined in \eqref{functional} is differentiable in $S$ and its derivative is given by
$$ \partial_S C\big(t,S,\xi \big) =  \frac{1}{2 \pi} \int_{b \in \mathbb R} \frac{a+ib}{S} \hat{g}(-b)L(a+ib,t,S,\xi)db. $$
Moreover, the function $C(t,S,.)$ is differentiable in the sense of Fr\'echet in $\xi$, with derivative such that for any $\zeta \in {\cal V}_{\alpha,\lambda}$,
$$ \partial_V C\big(t,S,\xi).\zeta =  \int_0^t \big(\frac{1}{2 \pi} \int_{b \in \mathbb R} \hat{g}(-b) L(a+ib,t,S,\xi) \chi(a+ib,t-s) db\big) \zeta(s) ds. $$
\end{prop}

\noindent We end this section by stating our result showing how one can build a hedging portfolio by trading the underlying and the forward variance curve.

\begin{theorem} \label{hedge} For any time $t \in [0,T]$, we have 
$$ C_t = C_0 + \int_0^t \partial_S C(T-u,S_u,\mathbb E[V_{.+u}|{\cal F}_u]) dS_u + \int_0^t \partial_V C(T-u,S_u,\mathbb E[V_{.+u}|{\cal F}_u]).(d\mathbb E[V_{.+u}|{\cal F}_u]), $$
where 
$$\partial_V C(T-u,S_u,\mathbb E[V_{.+u}|{\cal F}_u]).(d\mathbb E[V_{.+u}|{\cal F}_u])$$
denotes
$$  \int_0^{T-u} \big(\frac{1}{2 \pi} \int_{b \in \mathbb R} \hat{g}(-b) L(a+ib,T-u,S_u,\mathbb E[V_{.+u}|{\cal F}_u]) \chi(a+ib,T-u-s) db\big) d\mathbb E[V_{s+u}|{\cal F}_u] ds,$$
with $d\mathbb E[V_{x}|{\cal F}_u]$ the Ito differential at time $u$ of the martingale $M_u=\E[V_{x}|{\cal F}_u]$, $u\leq x$.
\end{theorem}
\begin{remark}
We actually also show that
$$ d\mathbb E[V_{s+u}|{\cal F}_u]  = \frac{1}{\lambda} f^{\alpha,\lambda}(s) \nu \sqrt{V_u}dB_u.$$
\end{remark}

\noindent The proof of Theorem \ref{hedge} is given in Section \ref{finalproof}. This result shows that in an idealistic setting where the underlying asset and the forward variance curve can be traded (in continuous time), perfect replication
can be obtained in generalized rough Heston models. Of course, in practice, this strategy will be discretized and one will use liquid variance swaps or European options instead of the forward variance curve.

\begin{remark}
It is interesting to remark that the price function $C(t,S,\xi)$ is solution of a Feynman-Kac type path-dependent partial differential equation. Let us define the following derivative according to time $t>0$:
$$ \partial_tC(t,S,\xi) = \underset{\varepsilon \rightarrow 0^+}{\lim} \frac{1}{\varepsilon} \big(C(t-\varepsilon,S,\xi_{\varepsilon+.})-C(t,S,\xi)\big).$$
We easily have that $L(a+ib,t,S,\xi)$ is solution of the following path-dependent PDE:
$$ 0 = \partial_t L + \frac{1}{2} (S \sqrt{\xi_0})^2 \partial_S^2 L + \frac{1}{2} (\nu \sqrt{\xi_0}) \partial_V^2 L.(\frac{1}{\lambda}f^{\alpha,\lambda},\frac{1}{\lambda}f^{\alpha,\lambda}) + \rho (S \sqrt{\xi_0}) (\nu \sqrt{\xi_0})  \partial_{S,V}^2 L.(\frac{1}{\lambda}f^{\alpha,\lambda}),$$
with the initial condition $L(a+ib,0,S,\xi) = S^{a+ib}$.\\

\noindent As in Proposition \ref{regularity}, we can show that $C$ is twice differentiable in $S$ and in $V$ (in the sense of Fr\'echet for $V$), and that $\partial_t C$ is well-defined. So we can deduce that $C$ satisfies the same path-dependent PDE:
$$ 0 = \partial_t C + \frac{1}{2} (S \sqrt{\xi_0})^2 \partial_S^2 C + \frac{1}{2} (\nu \sqrt{\xi_0}) \partial_V^2 C.(\frac{1}{\lambda}f^{\alpha,\lambda},\frac{1}{\lambda}f^{\alpha,\lambda}) + \rho (S \sqrt{\xi_0}) (\nu \sqrt{\xi_0})  \partial_{S,V}^2 C.(\frac{1}{\lambda}f^{\alpha,\lambda}),$$
with the initial condition $C(0,S,\xi) = (S - K)_+$.\\

\noindent Note that $ \partial_V^2 C.(\frac{1}{\lambda}f^{\alpha,\lambda},\frac{1}{\lambda}f^{\alpha,\lambda})$ (resp. $ \partial_{S,V}^2 C.(\frac{1}{\lambda}f^{\alpha,\lambda})$) is the second Fr\'echet derivative of $C$ (resp. the first Fr\'echet derivative of $\partial_S C$) applied on $(\frac{1}{\lambda}f^{\alpha,\lambda},\frac{1}{\lambda}f^{\alpha,\lambda})$ (resp. $\frac{1}{\lambda}f^{\alpha,\lambda}$) which is well-defined even though $\frac{1}{\lambda} f^{\alpha,\lambda}$ does not belong to the metric space ${\cal V}_{\alpha,\lambda}$.
\end{remark}

\section{Proofs}
\label{proofs}
The notion of fractional integrals and derivatives are heavily used in the proofs. Notations, definitions and useful results related to them are given in Appendix \ref{fracCal}.
\subsection{Proof of Theorem \ref{condLaw}}\label{proofLaw}
\paragraph{Finding the dynamic of $V_t^{t_0}$ conditional on ${\cal F}_{t_0}$}
Using stochastic Fubini theorem, we can show that $I^{1-\alpha}V$ is a semi-martingale and for $t>0$,
$$ (I^{1-\alpha}V)_t = V_0 \int_0^t \frac{s^{-\alpha}}{\Gamma(1-\alpha)} ds + \int_0^t \lambda (\theta^0(s) - V_s) ds + \int_0^t \nu \sqrt{V_s} dB_s .$$
Therefore,
$$\frac{1}{\Gamma(1-\alpha)} \int_0^{t+{t_0}} (t+{t_0}-u)^{-\alpha} V_u du $$
is equal to
$$ \frac{1}{\Gamma(1-\alpha)} \int_0^{{t_0}} ({t_0}-u)^{-\alpha} V_u du + V_0\int_{t_0}^{t+{t_0}} \frac{1}{\Gamma(1-\alpha)} u^{-\alpha} du  +\int_{t_0}^{t+{t_0}} \lambda (\theta^0(u) - V_u) du +  \int_{t_0}^{t+{t_0}} \nu \sqrt{V_u} dB_u.$$
Using a change of variable, this can be written
$$ \frac{1}{\Gamma(1-\alpha)} \int_0^{{t_0}} ({t_0}-u)^{-\alpha} V_u du + V_0\int_{0}^{t} \frac{1}{\Gamma(1-\alpha)} (t_0+u)^{-\alpha} du +\int_{0}^{t} \lambda (\theta^0(u+t_0) - V_u^{t_0}) du +  \int_{0}^{t} \nu \sqrt{V_u^{t_0}} dB_u^{t_0},$$
where $(B_t^{t_0})_{t \geq 0} = (B_{t+{t_0}}-B_{t_0})_{t\geq 0}$ is a Brownian motion independent of ${\cal F}_{t_0}$. Moreover, remarking that
$$  I^{1-\alpha} V^{t_0}_t = \frac{1}{\Gamma(1-\alpha)} \int_0^{t} (t-u)^{-\alpha} V_u^{t_0} du = \frac{1}{\Gamma(1-\alpha)} \int_{t_0}^{t+t_0} (t+t_0-u)^{-\alpha} V_u du  $$ is equal to 
$$  \frac{1}{\Gamma(1-\alpha)} \int_0^{t+t_0} (t+t_0-u)^{-\alpha} V_u du - \frac{1}{\Gamma(1-\alpha)} \int_0^{t_0} (t+t_0-u)^{-\alpha} V_u du$$
and that
$$ \frac{1}{\Gamma(1-\alpha)}\big((t_0-u)^{-\alpha} - (t+t_0-u)^{-\alpha}\big)= \frac{\alpha}{\Gamma(1-\alpha)} \int_0^t ({t_0}-u+v)^{-1-\alpha} dv,$$
we derive
\begin{align*} 
I^{1-\alpha} V^{t_0}_t  &=  \frac{\alpha}{\Gamma(1-\alpha)} \int_0^{{t_0}}  \int_0^t ({t_0}-u+v)^{-1-\alpha} dv V_u du + \int_{0}^{t} \frac{1}{\Gamma(1-\alpha)} (t_0+u)^{-\alpha} du V_0 \\
&+\int_{0}^{t} \lambda (\theta^0(u+t_0) - V_u^{t_0}) du +  \int_{0}^{t} \nu \sqrt{V_u^{t_0}} dB_u^{t_0}.
\end{align*}
This can be written as follows:
\begin{equation}\label{fracInt} 
V_{t_0} \frac{t^{1-\alpha}}{(1-\alpha)\Gamma(1-\alpha)} + \int_0^t \lambda(\theta^{t_0}(u) - V^{t_0}_u) du +   \int_0^{t} \nu \sqrt{V_u^{t_0}} dB_u^{t_0},  
\end{equation}
with $ (\theta^{t_0}(u))_{u \geq 0} $ a function which is ${\cal F}_{t_0}$ measurable and defined by 
$$ \theta^{t_0}(u) = \theta^0(t_0+u) + \frac{\alpha}{\lambda \Gamma(1-\alpha)} \int_0^{{t_0}} ({t_0}-v+u)^{-1-\alpha}  (V_v-V_{t_0}) dv  +  \frac{(u+{t_0})^{-\alpha}}{\lambda \Gamma(1-\alpha)}(V_0-V_{t_0}).$$

\paragraph{Properties of $\theta^{t_0}$} It is clear that $\theta^{t_0}$ is continuous on $\mathbb R_+^*$. Moreover it is easy to see that for any $u > 0$:
$$ \theta^{t_0}(u) = \theta^0(t_0+u) + \frac{\alpha}{\lambda \Gamma(1-\alpha)} \int_0^{t_0} ({t_0}-v+u)^{-1-\alpha} V_v  dv  + \frac{1}{\lambda \Gamma(1-\alpha)}(V_0(u+{t_0})^{-\alpha} -  V_{t_0} u^{-\alpha}).$$
Since $V$ is a non-negative process and $\theta^0$ satisfies \eqref{condition1}, we obtain that $\theta^{t_0}$ also satisfies \eqref{condition1}.
Finally, for fixed $\varepsilon>0$, $V$ being $\alpha-1/2-\varepsilon$ H\"older continuous, there exists for almost each $\omega \in \Omega $ a positive constant $c_\varepsilon(\omega)$ such that for any $x,y \in [0,t_0]$:\\
$$ |V_x - V_y| \leq c_\varepsilon(\omega) |x-y|^{\alpha-1/2-\varepsilon} .$$ 
Thus by integration by parts, we obtain for any $ u \in (0,t_0]$
\begin{align*}
| \int_0^{{t_0}} ({t_0}-v+u)^{-1-\alpha}  (V_v-V_{t_0}) dv| &\leq c_\varepsilon(\omega)  \int_0^{t_0}(t_0-v+u)^{-1-\alpha}(t_0-v)^{\alpha-1/2-\varepsilon}dv \\
&=  c_\varepsilon(\omega)  u^{-1/2-\varepsilon} \int_0^{t_0/u}(x+1)^{-1-\alpha}x^{\alpha-1/2-\varepsilon}dx \\
&\leq c_\varepsilon(\omega)  u^{-1/2-\varepsilon} \int_0^{\infty}(x+1)^{-1-\alpha}x^{\alpha-1/2-\varepsilon}dx.
\end{align*}
Thus $\theta^{t_0}$ satisfies Condition \eqref{condition2} almost surely. 

\paragraph{End of the proof}
We end the proof noting that from \eqref{fracInt} and stochastic Fubini Theorem we have that
$$\int_0^t V_s^{t_0} ds = \frac{1}{\Gamma(\alpha)}\int_0^t (t-s)^{\alpha-1} I^{1-\alpha}V^{t_0}_s ds $$
is equal to 
$$ V_{t_0} t +  \frac{1}{\Gamma(\alpha)} \int_0^t  \int_0^s (s-u)^{\alpha-1}  \lambda(\theta^{t_0}(u) - V^{t_0}_u) du ds +  \frac{1}{\Gamma(\alpha)} \int_0^{t} \int_0^s (s-u)^{\alpha-1} \nu \sqrt{V_u^{t_0}} dB_u^{t_0} ds.$$
Hence by differentiating the previous equality, we conclude that the dynamic of $(S^{t_0},V^{t_0})$ is given by
$$ S^{t_0}_t = S_{t_0} \exp\big( \int_0^t \sqrt{V^{t_0}_u} dW_u^{t_0} - \frac{1}{2} \int_0^t V_u^{t_0} du \big), $$
$$ V_t^{t_0} = V_{t_0} + \frac{1}{\Gamma(\alpha)}\int_0^t (t-u)^{\alpha-1} \lambda(\theta^{t_0}(u) - V_u^{t_0}) du + \frac{1}{\Gamma(\alpha)}\int_0^t (t-u)^{\alpha-1} \nu\sqrt{V_u^{t_0}} dB_u^{t_0} ,$$
where $(W_t^{t_0})_{t \geq 0} = (W_{t+{t_0}}-W_{t_0})_{t\geq0}$ is a Brownian motion independent of ${\cal F}_{t_0}$ and with correlation $\rho$ with $B^{t_0}$.

\subsection{Proof of Theorem \ref{moment1}}\label{proofMoment1}
We work here with the sequence of Hawkes processes $N^T$ defined in Assumption \ref{assump1Hawkes}. Recall that for $t\geq 0$, from Theorem \ref{ConvergenceTheorem}, $ \nu^2 T^{-2 \alpha} N_{tT}^T$, converges in law as $T$ goes to infinity to
$$ \int_0^t V_s ds ,$$
where $V$ is solution of the fractional stochastic differential equation \eqref{hestonGeneral}. 
A key step for the proof of Theorem \ref{moment1} is to show that for suitable  $a \in \mathbb R$,
\begin{equation} \label{uniformInteg} 
\mathbb E [\text{exp}(a\nu^2 T^{-2 \alpha} N_{tT}^T)]  \underset{T \rightarrow \infty}{\longrightarrow} \mathbb E [\text{exp}(a \int_0^t V_s ds)].
\end{equation}
Applying \eqref{recursiveCharac} in Appendix \ref{branching} on the Hawkes process $N^T$, we write
$$ \mathbb E [\text{exp}(a \nu^2 T^{-2 \alpha} N_{tT}^T)] = \text{exp}\big( \int_0^t \lambda \zeta^T(T(t-s)) g^T(a,s) ds \big), $$
with 
$$ g^T(a,t) =\nu^{-2} T^{\alpha} \big(\text{exp}(a\nu^2T^{-2 \alpha})\mathbb E [\text{exp}(a\nu^2T^{-2 \alpha} N^{f,T}_{tT})]-1\big),$$
where $N^{f,T}$ is the Hawkes process of children cluster (with migrant rate $\varphi^T$ and kernel $\varphi^T$), see Appendix \ref{branching} for details. Moreover from Lemma \ref{lemma2}, $\lambda \zeta^T(Ts)$ converges pointwise as $T$ goes to infinity to
$$ \lambda \theta^0(s) + \frac{V_0 s^{-\alpha}}{\Gamma(1-\alpha)}, \quad 0 < s \leq t .$$
Therefore, it is left to study the convergence of the function $g^T$.

\paragraph{Uniform boundedness of $g^T$}\label{boundness}
From now on $c$ denotes a positive constant that may vary from line to line.\\

\noindent 
From \eqref{condFinal} in Appendix \ref{momentHawkes}, for each $t>0$, 
\begin{equation} \label{conditionInfinity} 
g^T(a,t) < \infty 
\end{equation}
provided 
$$ a\nu^2T^{-2 \alpha} \leq  \int_0^{tT}\varphi^T -1 - \log(\int_0^{tT} \varphi^T). $$
Moreover note that from Appendix \ref{mittag},
$$ T^{2 \alpha} \big( \int_0^{tT}\varphi^T -1 - \log(\int_0^{tT} \varphi^T) \big) \underset{T \rightarrow \infty}{\longrightarrow} \frac{1}{2} (\lambda + \frac{\alpha t^{-\alpha}}{\Gamma(1-\alpha)})^2.$$
Thus Property \eqref{conditionInfinity} is satisfied for large enough $T>T_0(a,t,\lambda,\nu)$ and $a < a_0(t)$ with 
$$
a_0(t) = \frac{1}{2 \nu^2} (\lambda + \frac{\alpha t^{-\alpha}}{\Gamma(1-\alpha)})^2.$$ 
Furthermore, as $N^{f,T} \leq N^{\infty,T}$ (which is the Galton-Watson process defined in Appendix \ref{GW}), using \eqref{EaNinfty} in Appendix \ref{GW}, we obtain
$$ \mathbb E [\text{exp}(a\nu^2T^{-2 \alpha} N^{f,T}_{tT})]  \leq  \underset{n \geq 0}{\sum} \frac{\nu_T(t)^n e^{-\nu_T(t)(n+1)}}{n!} (n+1)^{n-1} e^{a\nu^2T^{-2 \alpha}n},$$
with $\nu_T(t) = \int_0^{tT} \varphi^T.$
It is also easy to see from \eqref{EaNinfty} (by taking $a = 0$ and $\nu=1$ in \eqref{EaNinfty}) that
$$ 1 =  \underset{n \geq 0}{\sum} \frac{e^{-(n+1)}}{n!} (n+1)^{n-1} .$$
Consequently, we obtain
\begin{align*}
g^T(a,t) &\leq \nu^{-2} T^{\alpha}  \underset{n \geq 0}{\sum} \frac{e^{-(n+1)}}{n!} (n+1)^{n-1} \big( \nu_T(t)^n e^{(1-\nu_T(t))(n+1)} e^{a\nu^2T^{-2 \alpha}(n+1) } - 1 \big)\\
&= \frac{1}{\nu^{2} \nu_T(t)}  T^{\alpha}  \underset{n \geq 0}{\sum} \frac{e^{-(n+1)}}{n!} (n+1)^{n-1} \big(e^{x_T(t)(n+1)} - \nu_T(t)\big),
\end{align*}
where
$$ x_T(t) = 1-\nu_T(t) + \log(\nu_T(t)) + a\nu^{2}T^{-2 \alpha},$$
which is non-positive for $T>T_0(a,t,\lambda,\nu)$.
Therefore
$$ g^T(a,t) \leq \frac{1}{\nu^2 \nu_T(t)} T^{\alpha} (1-\nu_T(t)).$$

\noindent Assume now $a \leq 0$, we use again $N^{f,T} \leq N^{\infty,T}$ and \eqref{EaNinfty} to get 
$$ \nu^{-2} T^{\alpha}  \underset{n \geq 0}{\sum} \frac{e^{-(n+1)}}{n!} (n+1)^{n-1} \big( e^{x_T(t)(n+1)} - 1 \big) \leq  g^T(a,t) \leq 0. $$
By Stirling formula,  
$$\frac{e^{-(n+1)}}{n!} (n+1)^{n-1} \underset{n \rightarrow \infty}{\sim}  \frac{1}{\sqrt{2 \pi (n+1)^3}}.$$
Thus,
$$ - c \frac{1}{\nu^2 \nu_T(t)}  T^{\alpha}  \underset{n \geq 0}{\sum} \frac{1}{(n+1)^{3/2}} (1 - e^{x_T(t)(n+1)})  \leq g^T(a,t) \leq 0.$$
We deduce from Lemma \ref{lemmtech} that
$$ -  \frac{c}{\nu^2 \nu_T(t)}   \sqrt{-T^{2\alpha} x_T(t)}  \leq g^T(a,t) \leq 0.$$

\noindent Therefore, for any $a < a_0(t)$:
$$  |g^T(a,t)| \leq c \frac{1}{\nu^2 \nu_T(t)}   \big( T^{\alpha}(1-\nu_T(t)) + \sqrt{-T^{2\alpha} x_T(t)}\big).$$
Finally, note that
$$ T^{\alpha}(1-\nu_T(t)) \rightarrow \frac{\alpha t^{-\alpha}}{\Gamma(1-\alpha)} $$
and
$$ T^{2\alpha}x_T(t) \rightarrow \nu^2(a_0(t)-a),$$
as $T$ goes to infinity. Eventually,
\begin{equation}  \label{IneqLimSup}
\underset{T \rightarrow \infty}{\limsup} |g^T(a,t)|  \leq c\nu^{-2}  \big( \frac{\alpha t^{-\alpha}}{\Gamma(1-\alpha)}+\nu\sqrt{(a_0(t)-a)} \big).
\end{equation}

\paragraph{Uniform convergence of $g^T$}\label{convergence}
We now fix $t_0>0$ and $a<a_0(t_0)$.
The function $t \rightarrow g^T(a,t)$ being monotone and such that $g^T(a,0) = 0$, we have that for any $0\leq t \leq t_0$ 
$$ |g^T(a,t)| \leq  |g^T(a,t_0)|. $$ 
Moreover, from the previous section, there exists $T_0(t_0,a,\lambda,\nu)>0$ such that 
$$ \underset{T \geq T_0 }{\sup} |g^T(a,t_0)| < \infty.$$
Hence, $g^T(a,.)$ is uniformly bounded in $0\leq t \leq t_0$ and $T\geq T_0$. We now assume that $T \geq T_0$.\\

\noindent Applying \eqref{recursiveCharac} in Appendix \ref{branching} on the Hawkes process $N^{f,T}$, we obtain that for any $t \in [0,t_0]$,
$$ \nu^2 T^{-\alpha}g^T(a,t) + 1= \exp\big(\nu^2T^{-2 \alpha} a+\nu^2 T^{1-\alpha}\int_0^t \varphi^T(Ts) g^T(a,t-s) ds \big) .$$
By taking the logarithm of the previous expression, we write
$$\nu^2 T^{-2 \alpha} a + \nu^2 T^{1-\alpha}\int_0^t \varphi^T(Ts) g^T(a,t-s) ds =  \nu^2 T^{-\alpha}g^T(a,t) - \frac{\nu^4}{2}  T^{-2\alpha} g^T(a,t)^2 - \varepsilon_1^T(t), $$
where $ |T^{3\alpha} \varepsilon_1^T|$ is uniformly bounded in $t \in [0,t_0]$ and $T\geq T_0$. Hence
$$ g^T(a,t) = T \int_0^t \varphi^T(Ts) g^T(a,t-s) ds + a T^{-\alpha}+  \frac{\nu^2}{2}T^{-\alpha} g^T(a,t)^2 +\frac{T^{\alpha}}{\nu^2}   \varepsilon_1^T(t) .$$
Thanks to Lemma \ref{hopf},
$$ g^T(a,t) =  a T^{1-\alpha} \int_0^t \psi^T(Ts) ds+  \frac{\nu^2}{2}T^{1-\alpha} \int_0^t \psi^T(Ts) g^T(a,t-s)^2 ds + \varepsilon_2^T(t), $$
where $$\varepsilon_2^T(t) = a T^{-\alpha}+  \frac{\nu^2}{2}T^{-\alpha} g^T(a,t)^2 +\frac{T^{\alpha}}{\nu^2}   \varepsilon_1^T(t) +\frac{T^{\alpha+1}}{\nu^2}  \int_0^t \psi^T(Ts) \varepsilon_1^T(t-s)ds$$
and $\displaystyle\psi^T= \sum_{k \geq 1} (\varphi^T)^{*k}$\footnote{Recall that $(\varphi^T)^{*1} = \varphi$ and $(\varphi^T)^{*k}(t) = \int_0^t \varphi(t-s).(\varphi^T)^{*k-1}(s) ds$.}.
Note that $ T^\alpha \varepsilon_2^T$ is uniformly bounded in $t  \in [0,t_0]$ and $T\geq T_0$. Recall also that using Laplace transform computations as in \cite{jaisson2016rough}, we get \begin{equation}\label{proppsi} \lambda T^{1-\alpha} \psi^T(Tt) = a_T f^{\alpha,\lambda}(t).\end{equation} Thus
 $$ g^T(a,t) =   \int_0^t \frac{1}{\lambda} f^{\alpha,\lambda}(t-s) (a + \frac{\nu^2}{2}g^T(a,s)^2 )ds + \varepsilon^T(t), $$
with $\varepsilon^T(t) = \varepsilon_2^T(t) - T^{-\alpha} \int_0^t f^{\alpha,\lambda}(t-s) (a + \frac{\nu^2}{2}g^T(a,s)^2 )ds $. As done in the proof of Proposition 6.5 in \cite{euch2016characteristic}, using that $T^{\alpha} \varepsilon^T$ and $g^T(a,.)$ are uniformly bounded in $t  \in [0,t_0]$ and $T \geq T_0$, together with Lemma \ref{UsefulIneq}, we deduce that  $g^T(a,.)$ is a Cauchy sequence on $C([0,t_0],\mathbb{R})$. Therefore it converges to a continuous function $g(a,.)$ solution of the following equation:
$$ g(a,t) =   \int_0^t \frac{1}{\lambda} f^{\alpha,\lambda}(t-s) (a + \frac{\nu^2}{2}g(a,s)^2 )ds.$$
By Lemma \ref{EDPFracLin}, it is equivalent to the fractional Riccati equation
$$ D^\alpha g(a,t) = a - \lambda g(a,t) + \frac{\nu^2}{2} g(a,t)^2, \quad I^{1-\alpha}g(a,0)= 0,$$
which admits a unique continuous solution (the uniqueness being an obvious corollary of Lemma \ref{UsefulIneq}).
Finally remark that from \eqref{IneqLimSup},
 $$|g(a,t)| \leq \frac{c}{\nu^2}  \big( \frac{\alpha t^{-\alpha}}{\Gamma(1-\alpha)}+ \nu \sqrt{a_0(t)-a} \big).$$
 
 \begin{remark} \label{inequalityRiccati1}Note that for $a\geq0$, $t \rightarrow g(-a,t)$ is non-increasing and since $g(-a,0) = 0$, we obtain the following inequality:
 $$ g(-a,t) =   \int_0^t \frac{1}{\lambda} f^{\alpha,\lambda}(t-s) (a + \frac{\nu^2}{2}g(-a,s)^2 )ds \leq \frac{1}{\lambda} F^{\alpha,\lambda}(t)(-a+ \frac{\nu^2}{2}g(-a,t)^2).$$
From this inequality, we get for $t>0$
$$ g(-a,t) \leq \frac{1-\sqrt{1+ \frac{2 \nu^2 a}{\lambda^2}F^{\alpha,\lambda}(t)^2}}{\frac{\nu^2}{\lambda}F^{\alpha,\lambda}(t)}.$$
\end{remark}

\paragraph{End of the proof}  We know that for any $t \in [0,t_0]$ and for fixed $a<a_0(t_0)$,
$$ \mathbb E [\text{exp}(a \nu^2 T^{-2 \alpha} N_{tT}^T)] = \exp\big( \int_0^t \lambda \zeta^T(T(t-s)) g^T(a,s) ds \big).$$
Then, from the uniform convergence of $g^T(a,.)$ to $g(a,.)$ together with Lemma \ref{lemma1}, Lemma \ref{lemma2} and the dominated convergence theorem, we obtain
$$ \mathbb E[\exp(a\nu^2T^{-2 \alpha}N_{tT}^T)] \rightarrow \exp\big( \int_0^t g(a,t-s)(\lambda \theta^0(s) + \frac{V_0 s^{-\alpha}}{\Gamma(1-\alpha)}) ds \big)$$
as $T$ goes to infinity. By Fatou lemma, we deduce
$$ \mathbb E[\exp\big(a \int_0^t V_s ds\big)] < \infty.$$

\noindent We end the proof by showing \eqref{uniformInteg}. The case $a \leq 0$ being obvious, we assume that $0<a<a_0(t_0)$. Let $\varepsilon>0$ such that $a(1+\varepsilon)<a_0(t_0)$. From the computations above, there exists $T_0(t,a,\lambda,\nu,\varepsilon)$ such that
$$ \underset{T\geq T_0}{\sup} \mathbb E[\exp(a(1+\varepsilon)\nu^2 T^{-2 \alpha}N_{tT}^T)] < \infty.$$
Therefore $\big(\text{exp}(a\nu^2T^{-2 \alpha}N_{tT}^T) \big)_{T\geq T_0} $ is uniformly integrable and we conclude that
$$ \mathbb E[\exp(a\nu^2T^{-2 \alpha}N_{tT}^T)] \rightarrow \mathbb E[\exp(a\int_0^t V_s ds)].$$
This ends the proof of Theorem \ref{moment1}.

\subsection{Proof of Theorem \ref{moment3}} \label{proofMoment3}
In this section, we place ourselves in the framework of generalized rough Heston models \eqref{hestonGeneral} and compute for $0\leq t\leq t_0$
$$ G(z,x,t) = \mathbb E [\text{exp}\big( z\int_0^t V_s ds + ix \int_0^t \sqrt{V_s} dB_s  \big)] ,$$
with $x \in \mathbb R$ and  $z \in \mathbb C$ such that $\Re (z) < a_0(t)$, ($a_0(t)$ is defined in Theorem \ref{moment1}).
It has been shown in the proof of Theorem \ref{moment1} that there exists $T_0>0$ such that
$$\text{exp}\big( z \nu^2 T^{-2 \alpha} N_{tT}^T+ ix  \nu T^{- \alpha} M_{tT}^T \big)$$
is uniformly integrable for fixed $t$ and $T\geq T_0$. We have that
\begin{equation}\label{eqrr}
\mathbb E [\text{exp}\big( z \nu^2 T^{-2 \alpha} N_{tT}^T+ ix  \nu T^{- \alpha} M_{tT}^T \big)]\end{equation} is equal to 
$$\mathbb E [\text{exp}\big( (z \nu^2 T^{-2 \alpha}+ix\nu T^{-\alpha})N_{tT}^T- ix\nu T^{-\alpha}\int_0^{tT}\int_0^s \varphi^T(s-u)dN_{u}^Tds- ix\nu T^{-\alpha}\mu_T \int_0^{tT}\zeta^T(s)ds\big)].$$ 
Let 
$$ f^T(t) = z \nu^2 T^{-2 \alpha} +ix  \nu T^{- \alpha}  - ix \nu T^{- \alpha}\int_0^{t}\varphi^T(s) ds .$$ Using Fubini theorem, we get that \eqref{eqrr} is also equal to
$$\mathbb E[\text{exp}\big( \int_0^{tT} f^T(tT-s) dN_s^T- ix\frac{\lambda}{\nu} \int_0^{t}\zeta^T(sT)ds\big)].$$
Hence we deduce from Lemma \ref{lemma1} and Lemma \ref{lemma2} that
\begin{align*} 
G(z,x,t) &= \underset{T \rightarrow \infty}{\lim} \mathbb E [\text{exp}\big( z \nu^2 T^{-2 \alpha} N_{tT}^T+ ix  \nu T^{- \alpha} M_{tT}^T \big)] \\
&= \text{exp}\big( -    \frac{ix}{\nu} \int_0^t \lambda \theta^0(s) + \frac{V_0 s^{-\alpha}}{\Gamma(1-\alpha)} ds \big)\underset{T \rightarrow \infty}{\lim} \mathbb E [\text{exp}\big( \int_0^{tT} f^T(tT-s) dN_s^T\big)].
\end{align*}

\paragraph{Passing to the limit}
Applying \eqref{exp2} in Appendix \ref{HawkesFunc} on the Hawkes process $N^T$ with the function $f^T$, we have that for large enough $T$,
$$ \text{exp}\big( \int_0^{tT} f^T(tT-s) dN_s^T\big) $$
is integrable and 
$$ \mathbb{E} [\text{exp}\big( \int_0^{tT} f^T(tT-s) dN_s^T\big)] = \text{exp}\big( \int_0^t \lambda \zeta^T(T(t-s)) k^T(z,x,s) ds \big)$$
where 
$$ k^T(z,x,t) = \frac{1}{\nu^2} T^\alpha \big(e^{f^T(tT)}\mathbb{E}[e^{\int_0^{tT} f^T(tT-u) dN_u^{f,T}}] -1\big).$$
Furthermore, from Lemma \ref{lemma2}, $ \lambda \zeta^T(Ts)$ converges pointwise as $T$ tends to infinity to
$$ \lambda \theta^0(s) + \frac{V_0 s^{-\alpha}}{\Gamma(1-\alpha)}, \quad s\leq t.$$
As in Section \ref{proofMoment1}, we show the uniform boundedness of $k^T$ and then its uniform convergence.

\paragraph{Uniform boundedness of $k^T$}
We start by noting that for $t \in [0,t_0]$, 
\begin{align*} 
|k^T(z,x,t)| &\leq \frac{1}{\nu^2} T^\alpha \big( \big|e^{f^T(tT)}\mathbb{E}[e^{\int_0^{tT} f^T(tT-u) dN_u^{f,T}}]  - e^{ i\Im[f^T(tT)]}\mathbb{E}[e^{\int_0^{tT} i \Im [f^T(tT-u)] dN_u^{f,T}}] \big| \\
&+ \big| e^{ i\Im[f^T(tT)]}\mathbb{E}[e^{\int_0^{tT} i \Im[f^T(tT-u)] dN_u^{f,T}}] -1 \big| \big).
\end{align*}
Using that $\Re[f^T] = \Re(z) \nu^2 T^{-2 \alpha}$ together with the following inequality
$$ |e^{f^T(tT)}\mathbb{E}[e^{\int_0^{tT} f^T(tT-u) dN_u^{f,T}}]  - e^{ i\Im[f^T(tT)]}\mathbb{E}[e^{\int_0^{tT} i \Im[f^T(tT-u)] dN_u^{f,T}}] \big| \leq  \big| e^{\Re[f^T(tT)]}\mathbb{E}[e^{\int_0^{tT} \Re[f^T(tT-u)] dN_u^{f,T}}]  - 1 \big|,$$
we derive
$$ |k^T(z,x,t)| \leq |k^T(\Re(z),0,t)| + |k^T(i\Im(z),x,t)|.$$
In Section \ref{proofMoment1}, we have already shown that $k^T(\Re(z),0,t)$ is uniformly bounded in $t \in [0,t_0]$, for large enough $T$. It is now left to show the uniform boundedness of $k^T(i\Im(z),x,t)$. So now we take $z = ia$ where $a\in \mathbb R$.
First, remark that $f^T$ becomes
$$ f^T(t) = ia\nu^2 T^{-2\alpha} + ix \nu T^{-\alpha}(1 - \int_0^t \varphi^T(s)ds ) = i(a \nu^2 + x \nu \lambda ) T^{-2\alpha} + ix \nu T^{-\alpha} \int_t^\infty \varphi^T(s)ds.$$
We write
$$ \tilde{X}^T_t = f^T(tT) + \int_0^{tT} f^T(tT-s) dN_s^{f,T} .$$
It is easy to see that
$$ |f^T(tT)| \leq |a|\nu^2 T^{-2\alpha}+|x| \nu T^{-\alpha}.$$
Furthermore,
$$ \mathbb{E}[\int_0^{tT} f^T(tT-s) dN_s^{f,T}] $$ is equal to
$$T^{-2\alpha} i(a\nu^2+\lambda x\nu) \int_0^{tT}\mathbb{E}[\lambda^{f,T}_s] ds + i x \nu T^{-\alpha} \int_0^{tT} \big(\int_{tT -s}^\infty \varphi^T(u) du \big) \mathbb{E}[\lambda^{f,T}_s] ds ,$$
where $\lambda^{f,T}$ is the intensity of the cluster of children Hawkes process $N^{f,T}$, see Appendix \ref{branching}. We recall its definition:
 $$\lambda^{f,T}_{u} = \varphi^T(u) + \int_0^u \varphi^T(u-s) dN_s^{f,T} . $$
Using Lemma \ref{hopf}, we know that
$$\lambda^{f,T}_{u} = \psi^T(u) + \int_0^u \psi^T(u-s) dM_s^{f,T} ,$$
where $M^{f,T} = N^{f,T} - \int_0^. \lambda^{f,T}_s ds $ is the martingale associated to $N^{f,T}$. Thanks to \eqref{proppsi}, we obtain 
$$ \mathbb{E}[\lambda^{f,T}_{tT}] = \frac{a_T f^{\alpha,\lambda}(t)}{\lambda} T^{\alpha -1} .$$
Therefore
$$\int_0^{tT}\mathbb{E}[\lambda^{f,T}_s] ds \leq \frac{F^{\alpha,\lambda}(t)}{\lambda} T^{\alpha} \leq \frac{F^{\alpha,\lambda}(t_0)}{\lambda} T^{\alpha} \leq c T^{\alpha}.$$
Moreover, using that $ y \in \mathbb R_+ \rightarrow y^{\alpha} \int_y^\infty \varphi^T$ is uniformly bounded in $y$ and $T$ and $I^{1-\alpha} f^{\alpha,\lambda} = \lambda(1-F^{\alpha,\lambda})$ (see Appendix \ref{mittag}), we obtain
$$T \int_0^{t} \int_{T(t -s)}^\infty \varphi^T(u) du \mathbb{E}[\lambda^{f,T}_{sT}] ds \leq \frac{c}{\lambda} \int_0^t (t-s)^{-\alpha} f^{\alpha,\lambda}(s) ds \leq c.$$
We deduce then that
$$ | \mathbb{E}[\tilde{X}^T_t]| \leq c T^{-\alpha}(|a|\nu^2+|x|\nu).$$

\noindent Using that there exists $c>0$ such that for any $y \in \mathbb R$,
$$ |e^{iy} - 1 -iy| \leq c y^2,$$
we get
$$ |\mathbb{E}[e^{\tilde{X}^T_t} - 1]| \leq c (| \mathbb{E}[\tilde{X}^T_t]| + \mathbb{E}[|\tilde{X}^T_t|^2]).$$
We have
\begin{align*}
\mathbb{E}[|\tilde{X}^T_t|^2] &\leq 2 |f^T(t)|^2 + 2 \mathbb{E}[|\int_0^{tT} f^T(tT-s) dN_s^{f,T}|^2]),\end{align*}
and 
\begin{align*}
\mathbb{E}[|\int_0^{tT} f^T(tT-s) dN_s^{f,T}|^2] &\leq 2( \mathbb{E}[|\int_0^{tT} f^T(tT-s) dM_s^{f,T}|^2]  + \mathbb{E}[|\int_0^{tT} f^T(tT-s) \lambda^{f,T}_s ds|^2] ).
\end{align*}
Since $\langle M^{f,T},M^{f,T}  \rangle = \int_0^. \lambda^{f,T}(s) ds $, we obtain 
$$  \mathbb{E}[|\int_0^{tT} f^T(tT-s) dM_s^{f,T}|^2]  =  \mathbb{E}[\int_0^{tT} |f^T(tT-s)|^2 \lambda_s^{f,T} ds] \leq c T^{-\alpha}(|a| \nu^2+ |x| \nu)^2.$$
Using Fubini theorem, we derive
$$ \int_0^{tT} f^T(tT-s) \lambda^{f,T}_s ds = \int_0^{tT} f^T(tT-s) \psi^T(s) ds  + \int_0^{tT}  \int_0^{tT-s}f^T(tT-s-u) \psi^T(u) du dM^{f,T}_s.$$
Therefore,
\begin{align*}
&\mathbb{E}[|\int_0^{tT} f^T(tT-s) \lambda^{f,T}_s ds|^2]\\
&\leq  2|\int_0^{tT} f^T(tT-s) \psi^T(s) ds|^2  +2\int_0^{tT}  |\int_0^{tT-s}f^T(tT-s-u) \psi^T(u) du|^2 \mathbb{E}[\lambda^{f,T}_s ]ds \\
&\leq c  (|a| \nu^2+ |x| \nu)^2 T^{-2 \alpha} (1+ \int_0^{tT}  \mathbb{E}[\lambda^{f,T}_s ]ds)\\
&\leq c  (|a| \nu^2+ |x| \nu)^2 T^{- \alpha}.
\end{align*}
We eventually deduce
$$ |k^T(ia,x,t)| \leq  \frac{c}{\nu^2}\big(c(a,x)+c(a,x)^2\big), \quad c(a,x) = \nu^2 |a| + \nu |x|.$$

\paragraph{End of the proof}
Using the same computations as in Section \ref{convergence}, we show that for fixed $z \in \mathbb C$ and $x \in \mathbb R$ such that $\Re(z) < a_0(t_0)$, $k^T(z,x,.)$ is a Cauchy sequence in $C([0,t_0],\mathbb C)$ and therefore converges uniformly to $k(z,x,.)$ solution of 
$$ k(z,x,t) = ix\frac{1}{\nu} + \int_0^t \frac{1}{\lambda} f^{\alpha,\lambda}(t-s) \big( z + \frac{\nu^2}{2}k(z,x,t)^2 \big) ds.$$
Therefore, we deduce 
$$ G(z,x,t) = \text{exp}\big(\int_0^t \xi(z,x,t-s) (\lambda \theta^0(s) + \frac{V_0 s^{-\alpha}}{\Gamma(1-\alpha)}) ds \big) $$
where $\xi(z,x,t) = k(z,x,t) -  ix/\nu$, which is solution of the following equation:
$$\xi(z,x,t) =   \int_0^t \frac{1}{\lambda} f^{\alpha,\lambda}(t-s) \big( z - \frac{x^2}{2} + ib \nu \xi(z,x,s) +\frac{\nu^2}{2}\xi(z,x,s)^2 \big) ds.$$
By Lemma \ref{EDPFracLin}, this is equivalent to the following fractional Riccati equation:
$$ D^{\alpha} \xi(z,x,t) = z - \frac{x^2}{2} + (ix\nu - \lambda) \xi(z,x,t) +\frac{\nu^2}{2}\xi(z,x,t)^2, \quad I^{1-\alpha}\xi(z,x,0) = 0. $$

\noindent We end this section with the following remarks which will be useful in the proof of Proposition \ref{regularity}.
\begin{remark} \label{inequalityRiccati2}
From the definition of $k^T$, 
$$ \Re(k^T(z,x,t)) \leq k^T(\Re(z),0,t).$$
Passing to the limit as $T$ goes to infinity, we get
$$ \Re(\xi(z,x,t)) \leq \xi(\Re(z),0,t) = g(\Re(z),t),$$ 
with $g$ defined in Theorem \ref{moment1}.
\end{remark}

\begin{remark} \label{UniformBound} From the proof of uniform boundedness of $k^T$ and using the inequality for $g$ in Theorem \ref{moment1}, we get that for any $t \in [0,t_0]$,
$$|\xi(z,x,t)| \leq c (1+\sqrt{|\Re(z)|} + \Im(z)^2 + x^2) ,$$
where $c$ is a positive constant, $x \in \mathbb R$ and $z \in \mathbb C$ such that $\Re(z) < a_0(t_0)$.
\end{remark}

\subsection{Proof of Proposition \ref{regularity}}\label{proofregu}
We fix $S>0$, $t>0$, $\xi \in {\cal V}_{\alpha,\lambda}^+$ and $a>1$ such that $\mathbb E[(S_t)^a] < \infty $.
Using the same computations as in the proof of Corollary \ref{moment5}, we get
$$L(a+ib,t,S,\xi) = \text{exp}\big((a+ib) \log(S)+ \int_0^{t} \frac{s^{-\alpha}}{\Gamma(1-\alpha)}h(a+ib,t-s) \theta_\xi(s)ds \big), $$
where $h$ is the unique continuous solution of the fractional Riccati equation in Corollary \ref{moment4}. Moreover, thanks to Remark \ref{inequalityRiccati2}, we have that for any $t>0$ and $b \in \mathbb R$,
$$ \Re\big(h(a+ib,s)\big) \leq q(b,s),\quad s \leq t , $$
where $q(b,.)$ is the unique continuous solution of the following fractional Riccati equation:
$$ D^\alpha q(b,s) = \frac{a^2-a}{2}-(1-\rho^2)\frac{b^2}{2} - (\lambda-\rho \nu a) q(b,s) + \frac{\nu^2}{2} q(b,s)^2,\quad s\leq t, \quad  I^{1-\alpha} q(b,0) = 0.$$
Note also that for large $|b|$, $\frac{a^2-a}{2}-(1-\rho^2)\frac{b^2}{2}$ is negative and therefore, using Remark \ref{inequalityRiccati1},
$$ q(b,s) \leq M(b,s)= \frac{1-\sqrt{1+\nu^2 \frac{(1-\rho^2)b^2-(a^2-a)}{(\lambda-\rho \nu a)^2} F^{\alpha,\lambda-\rho \nu a}(s)^2}}{\frac{\nu^2}{\lambda-\rho \nu a}F^{\alpha,\lambda-\rho \nu a}(s)},\quad s \leq t.$$
By dominated convergence theorem, we have
$$ \int_0^t  \frac{s^{-\alpha}}{\Gamma(1-\alpha)}M(b,t-s) \theta_\xi(s) ds \underset{b \rightarrow \infty}{\sim} -|b| \frac{\sqrt{1-\rho^2}}{\nu}  \int_0^t  \frac{s^{-\alpha}}{\Gamma(1-\alpha)} \theta_\xi(s)ds.$$
Consequently there exists $c(t,\xi)>0$ such that for any $b \in \mathbb{R} $,
$$ |L(a+ib,t,S,\xi)| \leq S^a \text{exp}\big( - c(t,\xi) (-1+|b|) \big).$$ 
Moreover, it is easy to see that for any $b \in \mathbb R$, $L(a+ib,t,.,\xi)$ is differentiable in $S$ and that
$$  \partial_S L(a+ib,t,S,\xi) =  \frac{a+ib}{S}L(a+ib,t,S,\xi) .$$
Using \eqref{parseval} together with the dominated convergence theorem, we conclude that $C$ is differentiable in the first variable $S$ and that
$$ \partial_S C(t,S,\xi) = \frac{1}{2\pi} \int_{b \in \mathbb R} \hat{g}(-b)\frac{a+ib}{S}L(a+ib,t,S,\xi) db.$$

\noindent Now let $\zeta \in {\cal V}_{\alpha,\lambda}$ and $\varepsilon_0>0$ such that $\theta_\xi(s) - \varepsilon_0 |\theta_\zeta(s)| > 0$ for any $s \in [0,t]$. We have that for any $\varepsilon\neq 0$, $\varepsilon\in (-\varepsilon_0,\varepsilon_0)$,
\begin{equation} \label{variation}\frac{1}{\varepsilon}|L(a+ib,t,S,\xi+\varepsilon \zeta)-L(a+ib,t,S,\xi)| \end{equation}
is equal to
\begin{align*}  S^a &  \text{exp}\big( \int_0^t \frac{s^{-\alpha}}{\Gamma(1-\alpha)} \Re(h(a+ib,t-s)) (\theta_\xi(s) - \varepsilon |\theta_\zeta(s)|) ds  \big)  \\
&\frac{1}{\varepsilon} \big| \text{exp}\big( \int_0^t \frac{s^{-\alpha}}{\Gamma(1-\alpha)} \varepsilon h(a+ib,t-s) \theta_\zeta(s)_- ds  \big)  - \text{exp}\big( \int_0^t \frac{s^{-\alpha}}{\Gamma(1-\alpha)} \varepsilon h(a+ib,t-s)  |\theta_\zeta(s)|ds  \big)  \big|  .
\end{align*}
Recall that for large $|b|$, $\Re\big(h(a+ib,s)\big)$ is non-positive for any $s \leq t$.
Since there exists $c>0$ such that for any $z,z' \in \mathbb C$ such that $\Re(z) \leq 0$ and $\Re(z') \leq 0$,
$$ |\text{exp}(z)-\text{exp}(z')| \leq c |z-z'|,$$
we conclude that \eqref{variation} is dominated by
$$ c S^a \big(\int_0^t \frac{s^{-\alpha}}{\Gamma(1-\alpha)} |h(a+ib,t-s)| |\theta_\zeta(s)| ds \big)  \text{exp}\big( \int_0^t \frac{s^{-\alpha}}{\Gamma(1-\alpha)} \Re(h(a+ib,t-s)) (\theta_\xi(s) - \varepsilon_0 |\theta_\zeta(s)|) ds  \big) .$$
Using the same arguments as previously, we get that there exists $c(t,\xi,\zeta,\varepsilon_0)>0$ such that
$$  \text{exp}\big( \int_0^t \frac{s^{-\alpha}}{\Gamma(1-\alpha)} \Re(h(a+ib,t-s)) (\theta_\xi(s) - \varepsilon_0 |\theta_\zeta(s)|) ds  \big) \leq \text{exp}\big(-c(t,\xi,\zeta,\varepsilon_0) (-1+|b|)\big).$$
From Remark \ref{UniformBound}, we know that there exists $c(t)>0$ such that for any $s\in [0,t]$ and $b\in\mathbb R$,
$$|h(a+ib,s)| \leq c(t) (1+b^2).$$
Moreover, note that
$$ \underset{\varepsilon \rightarrow 0}{\lim} \frac{1}{\varepsilon}L(a+ib,t,S,\xi+\varepsilon \zeta)-L(a+ib,t,S,\xi) $$
is equal to 
$$  L(a+ib,t,S,\xi) \int_0^t \chi(a+ib,t-s) \zeta_s ds. $$
Consequently, by the dominated convergence theorem, $C(t,S,.)$ is differentiable in $\xi$ in the direction of $\zeta$ in the Fr\'echet sense and
$$ \partial_V C(t,S,\xi).\zeta  = \frac{1}{2\pi} \int_{b \in \mathbb R} \hat{g}(-b)L(a+ib,t,S,\xi) \big( \int_0^t \chi(a+ib,t-s) \zeta_s ds\big)  db  .$$

\subsection{Proof of Theorem \ref{hedge}}\label{finalproof}
We first show that
$$ \int_0^{T-t} \chi(a+ib,s) \mathbb E[V_{T-s}|{\cal F}_t] ds$$
is equal to
\begin{equation} \label{ipp2}
\int_0^T \chi(a+ib,s) \mathbb E[V_{T-s}] ds  -\int_0^t \chi(a+ib,T-s) V_s ds +  \int_0^t h(a+ib,T-s) \nu \sqrt{V_s} dB_s.
\end{equation}
Recall that from Equation \eqref{dynV} we get 
$$ V_s = \mathbb E[V_s] + \int_0^s \frac{1}{\lambda} f^{\alpha,\lambda}(s-u) \nu \sqrt{V_u} dB_u.$$ This
together with stochastic Fubini theorem give
$$ \int_0^t \chi(a+ib,T-s) V_s ds =  \int_0^t \chi(a+ib,T-s) \mathbb E[V_s] ds + \int_0^t \big( \int_0^{t-u} \frac{1}{\lambda} f^{\alpha,\lambda}(s) \chi(a+ib,T-u-s) ds \big)  \nu \sqrt{V_u} dB_u.$$
We also have that for $s \in [0,T-t]$, 
\begin{equation}\label{forvarcurve}
\mathbb E [V_{T-s}|{\cal F}_t] =  \mathbb E[V_{T-s}] + \int_0^t \frac{1}{\lambda} f^{\alpha,\lambda}(T-s-u) \nu \sqrt{V_u} dB_u.
\end{equation}
Then similarly,
$$ \int_0^{T-t} \chi(a+ib,s) \mathbb E[V_{T-s}|{\cal F}_t]  ds $$
is equal to
$$ \int_0^{T-t} \chi(a+ib,s) \mathbb E[V_{T-s}] ds + \int_0^t  \big(\int_0^{T-t} \frac{1}{\lambda} f^{\alpha,\lambda}(T-s-u) \chi(a+ib,s) ds \big) \nu \sqrt{V_u} dB_u.$$
This can also be written
$$ \int_t^{T} \chi(a+ib,T-s) \mathbb E[V_{s}] ds + \int_0^t  \big(\int_{t-u}^{T-u} \frac{1}{\lambda} f^{\alpha,\lambda}(s) \chi(a+ib,T-u-s) ds \big) \nu \sqrt{V_u} dB_u.$$
Finally we obtain that
$$  \int_0^t \chi(a+ib,T-s) V_s ds +  \int_0^{T-t} \chi(a+ib,s) \mathbb E[V_{T-s}|{\cal F}_t]  ds $$
is equal to 
$$ \int_0^{T} \chi(a+ib,T-s) \mathbb E[V_{s}] ds + \int_0^t  \big(\int_{0}^{T-u} \frac{1}{\lambda} f^{\alpha,\lambda}(s) \chi(a+ib,T-u-s) ds \big) \nu \sqrt{V_u} dB_u.$$
Thus \eqref{ipp2} is directly deduced from the last relation and \eqref{hchi}. 
Now using \eqref{ipp2} together with Ito formula, we derive
$$ P^T_t(a+ib) =P^T_0(a+ib) + \int_0^t (a+ib) P^T_s(a+ib) \sqrt{V_s} dW_s + \int_0^t P^T_s(a+ib) h(a+ib,T-s) \nu \sqrt{V_s} dB_s.$$
Then by \eqref{parseval} together with stochastic Fubini theorem and Proposition \ref{regularity}, we get
$$ C_t = C_0 + \int_0^t \partial_S C(T-u,S_u,\mathbb E[V_{.+u}|{\cal F}_u]) dS_u + \frac{1}{2\pi} \int_0^t \big( \int_{b \in \mathbb R}\hat{g}(-b) P^T_u(a+ib) h(a+ib,T-u) db\big) \nu \sqrt{V_u} dB_u.$$
Furthermore, using again \eqref{hchi} together with Fubini theorem, we obtain that 
$$ \frac{1}{2\pi} \int_0^t \big( \int_{b \in \mathbb R}\hat{g}(-b) P^T_u(a+ib) h(a+ib,T-u) db\big) \nu \sqrt{V_u} dB_u $$
is equal to
$$  \int_0^t \big( \frac{1}{2\pi} \int_0^{T-u}  \int_{b \in \mathbb R}\hat{g}(-b) P^T_u(a+ib) \chi(a+ib,T-u-s) db ds\big) \frac{1}{\lambda} f^{\alpha,\lambda}(s) \nu \sqrt{V_u} dB_u.$$
This last quantity can be expressed in term of the forward variance curve thanks to \eqref{forvarcurve}.

\section*{Acknowledgments}

We thank Jim Gatheral for many interesting discussions.

\section*{Appendix}

\appendix

\section{Fractional calculus}\label{fracCal}
We define the fractional integral of order $r \in (0,1]$ of a function $f$ as
$$
I^r f(t) = \frac{1}{\Gamma(r)} \int_0^t (t-s)^{r-1} f(s) ds,  
$$
whenever the integral exists, and its the fractional derivative of order $r \in [0,1)$ as
$$D^r f(t) = \frac{1}{\Gamma(1-r)} \frac{d}{dt} \int_0^t (t-s)^{-r} f(s) ds,  
$$
whenever it exists.\\

\noindent We gather in this section some useful technical results related to fractional calculus.
\subsection{Mittag-Leffler functions}
\label{mittag}
Let $(\alpha,\beta) \in (\mathbb{R}_+^*)^2$. The Mittag-Leffler function $E_{\alpha,\beta}$ is defined and for $z \in \mathbb{C}$ by
$$ E_{\alpha,\beta}(z) = \sum_{n \geq 0} \frac{z^n}{\Gamma(\alpha n + \beta)}.$$ 
For $(\alpha,\lambda) \in  (0,1)\times \mathbb{R}_+$ we also define 
$$ f^{\alpha,\lambda}(t) = \lambda t^{\alpha - 1} E_{\alpha,\alpha}(-\lambda t^\alpha),~t>0, $$
$$ F^{\alpha,\lambda}(t) = \int_0^t f^{\alpha,\lambda}(s) ds,~t \geq 0.$$
The function $f^{\alpha,\lambda}$ is a density function on $\mathbb{R}_+$ called Mittag-Leffler density function.
The following properties of $f^{\alpha,\lambda}$ and $F^{\alpha,\lambda}$ can be found in \cite{haubold2011mittag,mainardisome,mathai2008special}.
We have
$$ f^{\alpha,\lambda}(t) \underset{t \rightarrow 0^+}{\sim} \frac{\lambda}{\Gamma(\alpha)} t^{\alpha-1},~~f^{\alpha,\lambda}(t) \underset{t \rightarrow \infty}{\sim} \frac{\alpha}{\lambda \Gamma(1-\alpha)} t^{-(\alpha+1)}$$
and
$$ F^{\alpha,\lambda}(t) = 1 - E_{\alpha,1}(-\lambda t^\alpha),~F^{\alpha,\lambda}(t) \underset{t \rightarrow 0^+}{\sim} \frac{\lambda}{\Gamma(\alpha+1)} t^{\alpha},~ 1-F^{\alpha,\lambda}(t) \underset{t \rightarrow \infty}{\sim} \frac{1}{\lambda \Gamma(1-\alpha)} t^{-\alpha}.$$
Note also that from obvious computations, we get $I^{1-\alpha} f^{\alpha,\lambda}= \lambda (1-F^{\alpha,\lambda}).$ Finally, for $\alpha \in (1/2,1)$, $f^{\alpha,\lambda}$ is square-integrable and its Laplace transform is given for $z \geq 0$ by 
$$\hat{f}^{\alpha,\lambda}(z)= \int_0^\infty f_{\alpha,\lambda}(s) e^{-zs} ds = \frac{\lambda}{\lambda + z^\alpha}.$$

\subsection{Wiener-Hopf equations}
The following result enables us to solve Wiener-Hopf type equations, see for example \cite{bacry2013some} for details.
\begin{lemma}
\label{hopf}
Let $g$ be a measurable locally bounded function from $\mathbb{R}$ to $\mathbb{R}^d$ and $\phi : \mathbb{R}_+ \rightarrow \cal{M}^{\textbf{d}}(\mathbb{R}) $ be a matrix-valued function with integrable components such that the spectral radius of $\int_0^\infty \phi(s) ds$ is strictly smaller than 1. Then there exists a unique
locally bounded function $f$ from $\mathbb{R}$ to $\mathbb{R}^d$ solution of
$$f(t) = g(t) + \int_0^t \phi(t-s). f(s) ds,~t \geq 0$$
given by
$$ f(t) = g(t) + \int_0^t \psi(t-s). g(s) ds,~t \geq 0,$$
where $\displaystyle\psi= \sum_{k \geq 1} \phi^{*k}$\footnote{Recall that 
$\phi^{*1} = \phi$ and $\phi^{*k}(t) = \int_0^t \phi(t-s).\phi^{*k-1}(s) ds$.}.
\end{lemma}

\subsection{\textsc{Fractional differential equations}}
We now give some useful results about fractional differential equations. The next lemma can be found in \cite{samko1993fractional}.
\begin{lemma}
\label{EDPFracLin} 
Let $h$ be a continuous function from $[0,1]$ to $\mathbb{R}$, $\alpha \in (0,1]$ and $\lambda\in\mathbb{R}$. There is a unique solution to the equation
\begin{equation*}
D^\alpha y(t) = \lambda y(t) + h(t),~~y(0) = 0
\end{equation*}
given by
$$ y(t) = \int_0^t (t-s)^{\alpha-1} E_{\alpha,\alpha}\big(\lambda(t-s)^\alpha\big) h(s) ds.$$
\end{lemma}
\noindent We also have the following result whose proof can be found in \cite{euch2016characteristic}.
\begin{lemma}
\label{UsefulIneq}
Let $h$ be a non-negative continuous function from $[0,1]$ to $\mathbb{R}$ such that for any $t \in[0,1]$,
$$ h(t) \leq \varepsilon + C \int_0^t f^{\alpha,\lambda}(t-s) h(s) ds,$$
for some $\varepsilon\geq 0$ and $C \geq 0$. Then for any $t \in[0,1]$, 
$$ h(t) \leq C' \varepsilon,$$
with $$C' = 1+ C\lambda \int_0^{1} s^{\alpha-1} E_{\alpha,\alpha}\big(\lambda(C-1)s^\alpha\big) ds>0.$$
In particular, if $\varepsilon =0$ then $h = 0$.
\end{lemma}

\subsection{Further results}

\begin{lemma}\label{lemma1}
There exists a positive constant $c$ such that for any $T>1/{\lambda^{-1/\alpha}}$ and $t\in (0,1)$:
$$ \zeta^T(tT) \leq c(1 + \frac{t^{-\alpha}}{\Gamma(1-\alpha)}).$$
\end{lemma}
\noindent \textsc{Proof of Lemma \ref{lemma1}:}\\ 

\noindent Note that by Remark \ref{mu}, we have 
$$ \zeta^T(tT) =  \int_0^{tT}   \varphi^T(tT-u) \theta^0(u/T) du + V_0  \big(\frac{T^\alpha}{\lambda}  \int_{tT}^\infty \varphi(s) ds  + \lambda T^{-\alpha} \int_0^{tT} \varphi(s) ds \big).$$
Thanks to Appendix \ref{mittag}, we have that for each $t\in (0,1]$:
$$ T^{\alpha} \int_{tT}^{\infty} \varphi \leq c t^{-\alpha}, $$
Moreover by using condition \eqref{condition2} and the fact that $\alpha > 1/2$, we write that for each $t\in (0,1]$:
$$ \theta^0(t) \leq c t^{-\alpha} .$$
Thus:
$$ \zeta^T(tT) \leq c \int_0^{tT}   \varphi(Tt-u) (u/T)^{-\alpha} du + c (1+t^{-\alpha}) . $$
Using Appendix \ref{mittag}, we obtain 
$$ \int_0^{tT}   \varphi(Tt-u) (u/T)^{-\alpha} du  = \Gamma(1-\alpha) T^{\alpha} \int_{tT}^\infty \varphi  \leq c t^{-\alpha},$$
which ends the proof.
\qed

\begin{lemma}\label{lemma2} For each $t \in (0,1]$, as $T$ tends to infinity, $\zeta^T(tT)$ defined by Assumption \ref{assump1Hawkes} converges to
$$  V_0 \frac{t^{-\alpha}}{\lambda \Gamma(1-\alpha)} + \theta^{0}(t).$$
\end{lemma}
\noindent\textsc{Proof of Lemma \ref{lemma2}:}\\

\noindent Let $t>0$. We have
$$ \zeta^T(tT) = a_T \int_0^t T \varphi(T(t-s)) \theta^0(s) ds + V_0    \big(\frac{T^\alpha}{\lambda}  \int_{tT}^\infty \varphi(s) ds  + \lambda T^{-\alpha} \int_0^{tT} \varphi(s) ds \big).$$
Moreover, from Appendix \ref{mittag},
$$ V_0   \big(\frac{T^\alpha}{\lambda}  \int_{tT}^\infty \varphi(s) ds  + \lambda T^{-\alpha} \int_0^{tT} \varphi(s) ds \big) $$
converges to
$$ V_0 \frac{t^{-\alpha}}{\lambda \Gamma(1-\alpha)}. $$
Moreover, since $\theta^0$ is continuous in $t$, for any $\varepsilon>0$ there exists $\eta>0$ such that for any $s \in [t-\eta,t]$,
$$ |\theta^0(s) - \theta^0(t)| \leq \varepsilon.$$
Hence from Appendix \ref{mittag} together with the fact that $\varphi$ is non-increasing, we obtain
\begin{align*}
\big|\int_0^t T \varphi\big(T(t-s)\big)\big(\theta^0(s)-\theta^0(t)\big) ds\big| &\leq \varepsilon \int_0^{T\eta}\varphi +  \int_0^{t-\eta} T \varphi\big(T(t-s)\big) (|\theta^0(t)|+|\theta^0(s)|) ds\\
&\leq \varepsilon + T \varphi(T\eta) \int_0^{t} ( |\theta^0(t)|+|\theta^0(s)|) ds\leq 2 \varepsilon
\end{align*}
for large enough $T$. Thus $ \int_0^t T \varphi(T(t-s)) \theta^0(s) ds$ converges to $\theta^0(t)$.\\
\qed

\begin{lemma}\label{lemma3} If $\theta^0 : (0,1] \rightarrow \mathbb{R}$ satisfies Condition \eqref{condition2}, then for any $0<\varepsilon<\alpha-1/2$, 
$$ t \rightarrow \int_0^t f^{\alpha,\lambda}(t-s) \theta^0(s) ds $$
has H\"older smoothness $\alpha - 1/2 - \varepsilon$ on $[0,1]$.
\end{lemma}
\noindent \textsc{Proof of Lemma \ref{lemma3}:}\\

\noindent Using Proposition A.2 in \cite{jaisson2016rough}, we obtain that for any $\eta \in (0,\alpha)$,
$$  \int_0^t f^{\alpha,\lambda}(t-s) \theta^0(s) ds = \int_0^t D^{\eta}f^{\alpha,\lambda}(t-s) I^{\eta}\theta^0(s).$$ Taking $\eta = 1/2 + \varepsilon$, we have that $I^{\eta}\theta^0$ is a bounded function. Then, using Proposition A.3 in \cite{jaisson2016rough}, we obtain that our function has H\"older regularity equal to $\alpha - \eta = \alpha - 1/2 - \varepsilon$.\\
\qed

\noindent Let $x \geq 0$. We define
$$ S(x) = \underset{n\geq 0}{\sum} \frac{1}{(n+1)^{3/2}} (1 - e^{-x(n+1)}).$$ We have the following lemma. 
\begin{lemma}\label{lemmtech} There exists $c>0$ such that for any $x \geq 0$:
$$ S(x) \leq c \sqrt{x}.$$ 
\end{lemma}

\noindent\textsc{Proof of Lemma \ref{lemmtech}}:\\
We have
$$ S(x) =  \underset{n\geq 0}{\sum} \frac{1}{(n+1)^{3/2}} (1 - e^{-x}) \underset{0 \leq k\leq n}{\sum} e^{-k x}.$$
This can be rewritten
$$ S(x) =  (1 - e^{-x})  \underset{k\geq 0}{\sum} \xi_k e^{-k x},$$
with $\xi_k = \underset{n \geq k }{\sum}  \frac{1}{(n+1)^{3/2}}$, which is equivalent to $ 2 / \sqrt{k+1} $ as $k$ tends to infinity. Thus there exists $c>0$ such that for any $x\geq 0$:
$$ S(x) \leq  c(1 - e^{-x})  \underset{k\geq 0}{\sum} \frac{1}{\sqrt{k+1}} e^{-(k+1) x}.$$
We conclude using that
$$ \underset{k\geq 0}{\sum} \frac{1}{\sqrt{k+1}} e^{-(k+1) x}  \leq \underset{k\geq 0}{\sum}  \int_k^{k+1}\frac{1}{\sqrt{y}} e^{-y x} dy = \frac{\Gamma(1/2)}{\sqrt{x}}$$
together with the fact that 
$$ 1 - e^{-x} \leq c x .$$
\qed

\section{Martingale property of the price in the generalized rough Heston model}
\begin{prop}\label{martingale} The process $S$ defined by the generalized rough Heston model in Definition \ref{GeneralRoughHeston} is a $\mathbb F$-martingale. 
\end{prop}

\noindent\textsc{Proof of Proposition \ref{martingale}}:\\
Let $t_0>0$ such that $1/2 < a_0(t_0) $. Thanks to Theorem \ref{moment1}, Novikov's criterion holds: 
$$ \mathbb{E}[\text{exp}(\frac{1}{2} \int_0^{t_0} V_s ds)]  < \infty .$$
Therefore $(S_u)_{0\leq u \leq t_0}$ is a martingale and $\mathbb E [S_{t_0}] = S_0$.\\

\noindent Now, assume that for a given $n \in \mathbb N$,  $\mathbb E [S_{n t_0}] = S_0$. Recall that conditional on ${\cal F}_{n t_0}$, the law of $(S_{t}^{nt_0},V_{t}^{nt_0})_{t \geq 0}=(S_{t+nt_0},V_{t+nt_0})_{t \geq 0}$ is still that of a rough Heston model with the following dynamic:
$$ dS_t^{nt_0} = S_t^{nt_0} \sqrt{V_t^{nt_0}} dW_t^{nt_0} $$
$$ V_t^{nt_0} = V_{nt_0} +\frac{1}{\Gamma(\alpha)} \int_0^t (t-u)^{\alpha - 1} \lambda (\theta^{nt_0}(u) - V_u^{nt_0})  du  + \frac{1}{\Gamma(\alpha)} \int_0^t (t-u)^{\alpha - 1}  \nu \sqrt{V_u^{nt_0}}  dB_u^{nt_0},
$$
where $\theta^{nt_0}$ is a ${\cal F}_{nt_0}$-measurable function satisfying almost surely Conditions \eqref{condition1} and \eqref{condition2}, and $(W^{nt_0},B^{nt_0}) = (W_{.+nt_0}-W_{nt_0},B_{.+nt_0}-B_{nt_0}) $ is a Brownian motion independent of ${\cal F}_{nt_0}$. Since $1/2 < a_0(t_0) $, we have again Novikov's criterion
$$ \mathbb{E}[\text{exp}(\frac{1}{2} \int_0^{t_0} V_s^{nt_0} ds)|{\cal F}_{nt_0}]  < \infty.$$
Therefore $\mathbb{E} [S_{t_0}^{nt_0}|{\cal F}_{nt_0}] = S_{nt_0}$ and so
$$ \mathbb{E} [S_{(n+1)t_0}] = \mathbb{E} [S_{nt_0}] = S_0 .$$
Consequently, for any $n \in \mathbb N$,
$$\mathbb{E}[S_{nt_0}] = S_0,$$
which ends the proof.
\qed

\section{Moments properties for Hawkes processes}\label{momentHawkes}
Here we consider a one-dimensional Hawkes process $N$ with intensity
$$ \lambda_t = \mu(t) + \int_0^t \varphi(t-s) dN_s,$$
such that $\mu,~\varphi : \mathbb R_+ \rightarrow \mathbb R_+$ are locally integrable and $\int_0^\infty \varphi < 1$.
We are interested in a sufficient condition on $a > 0$ so that
\begin{equation} \label{cond1}
 \mathbb{E}[e^{aN_t}] < \infty.
 \end{equation}
 We will show that \eqref{cond1} holds provided 
 \begin{equation} \label{condFinal}
a \leq \int_0^t\varphi -1 - \log(\int_0^t \varphi).
\end{equation}
To do so, we recall the branching structure of Hawkes processes.

\subsection{Branching structure of Hawkes processes}\label{branching}
We recall that the Hawkes process $N$ can be viewed as a population process in which migrants arrive according to a non-homogenous Poisson process $N^0$ with intensity $\mu$. Each migrant gives birth to children according to a non-homogenous Poisson process with intensity $\varphi$ and each child also gives birth to children according to non-homogenous Poisson process with the same intensity and so on.\\

\noindent Therefore, it is easy to see that the cluster of children created by a migrant has the law of a Hawkes process $N^f$ with the same kernel function $\varphi$ but with migrant rate $\varphi$. So, the intensity of $N^f$ is given by :
$$ \lambda_t^f = \varphi(t) + \int_0^t \varphi(t-s) dN_s^f.$$
\noindent Using the branching structure of the Hawkes process, we can see that we easily derive the following equality in law:
$$ N_t = N_t^0 + \underset{1\leq k \leq N^0_t}{\sum} N_{t-T_k}^{f,k} $$
where the $(T_k)_{k \geq 1}$ are the arrival times of the migrants and $(N^{f,k})_{k \geq 1}$ are independent copies of $N^f$, independent of $N^0$. Then, we can show that for $a \geq 0$,
\begin{equation}\label{recursiveCharac} 
\mathbb E [e^{aN_t}] = \text{exp}\big( \int_0^t \mu(t-s) (e^a\mathbb E [e^{aN_s^f}] -1) ds \big), 
\end{equation}
see \cite{euch2016characteristic}. This is smaller than
$$ \text{exp}\big( \int_0^t \mu(s) ds  (e^a\mathbb E [e^{aN_t^f}] -1) \big).$$
Consequently, a sufficient condition to obtain \eqref{cond1} is
\begin{equation} \label{cond2}
 \mathbb{E}[e^{aN_t^f}] < \infty.
 \end{equation}
 
\subsection{Galton-Watson structure and exponential moments}\label{GW}
Let us consider now the Hawkes process $N^f$. Using the population interpretation given in the previous section on this process, $N_t^f$ is the number of migrants and children arrived up to time $t$. Let $t>0$. We define the process $N^{\infty}$ from $N^f$ as follows. 
\begin{itemize}
\item We consider $N^{(0)}_t$ the number of migrants arrived up to time $t$, which is a Poisson variable with parameter $\nu = \int_0^t \varphi$.
\item For each migrant arrived at time $T_k < t$, we consider the number of children of first generation made by the migrant during a period of time $t$, which is also a Poisson variable with parameter $\nu$, independent of $N_t^{(0)}$. We denote by $X_t^1$ the set of all those children and $N_t^{(1)} = \#(X_t^1)$ their total number.
\item For each child of $n^{th}$ generation of the set $X_t^n$, we consider the number of its children that are made during a period of time $t$, which is also a Poisson variable with parameter $\nu$, independent of the previous generations. We denote $X_t^{n+1}$ the set of all those children and $N_t^{(n+1)} = \#(X_t^{n+1})$ their total number.
\end{itemize}
It is clear that $X_t = \underset{n \geq 0}{\bigcup} X^n_t$ contains all the individuals of the Hawkes process $N^f$ arrived up to time $t$. So,
$$ N^{\infty}_t =  \#(X_t) = \underset{n \geq 0}{\sum} N^{(n)}_t \geq N_t^f.$$
Thus a sufficient condition to obtain \eqref{cond2} is
\begin{equation} \label{cond3}
 \mathbb{E}[e^{aN^\infty_t}] < \infty .
\end{equation}
Now remark that $(N^{(n)}_t)_{n \geq 0}$ is a Galton-Watson process. Indeed,
$$ N^{(n+1)}_t = \underset{ 1 \leq k \leq N^{(n)}_t }{\sum} \xi_{k,n+1}; \quad n \geq 0 .$$
where $(\xi_{k,n})_{k,n \geq 1}$ are i.i.d Poisson random variables with parameter $\nu$, independent of the $N^{(k)}_t$. \\

\noindent We classically have, see for example \cite{dwass1969total},
$$ \mathbb P [N^\infty_t = n] = \frac{\nu^n e^{-\nu(n+1)}}{n!} (n+1)^{n-1}.$$
Consequently,
\begin{equation} \label{EaNinfty} 
\mathbb{E}[e^{aN^{\infty}_t}] = \underset{n \geq 0}{\sum} \frac{\nu^n e^{-\nu(n+1)}}{n!} (n+1)^{n-1} e^{an}.
\end{equation}
Using Stirling formula, we get
$$ \frac{\nu^n e^{-\nu(n+1)}}{n!} (n+1)^{n-1} e^{an} \underset{n \rightarrow \infty}{\sim} \frac{(\nu e^{1-\nu+a})^n}{\sqrt{2 \pi n^3}} e^{1-\nu}$$
Hence \eqref{cond3} holds if and only if 
$$ \nu e^{1-\nu+a} \leq 1,$$ 
which is equivalent to :
$$ a \leq \int_0^t\varphi -1 - \log(\int_0^t \varphi).$$

\subsection{A useful equality}\label{HawkesFunc}
Let us consider $g: \mathbb R_+ \rightarrow \mathbb R$ continuous and $a \in \mathbb R$ satisfying \eqref{condFinal}.
We know that $\text{exp}(\int_0^t f(t-s) dN_s)$ is integrable, where $f = a + ig$. Using the branching structure of Hawkes processes presented in Appendix \ref{branching}, we deduce the following equality in law:
$$ \int_0^t f(t-s) dN_s = \int_0^t f(t-s) dN_s^0 + \underset{1\leq k \leq N^0_t}{\sum} \int_0^{t-T_k} f(t-T_k-s) dN_s^{f,k}.  $$
Therefore, we can show that 
 \begin{equation} \label{exp2}
 \mathbb{E}[\text{exp}(\int_0^t f(t-s) dN_s)] = \text{exp}\big( \int_0^t \mu(t-s)(e^{f(s)}\mathbb{E}[e^{\int_0^s f(s-u) dN_u^f}] -1) ds\big).
 \end{equation}

\bibliographystyle{abbrv}
\bibliography{BibEER3_final}

\end{document}